\RequirePackage{etoolbox}
\csdef{input@path}{{style/}{graphics/}}
\documentclass[MSNbibl,nameyear,dvips]{arxstspdf}
\usepackage{url,breakurl}
\usepackage{graphicx}
\usepackage{flushend}
\usepackage{stfloats}

\volume{29}
\issue{4}
\pubyear{2014}
\firstpage{459}
\lastpage{484}
\doi{10.1214/14-STS505} 
\docsubty{FLA}

\makeatletter
\fnbelowfloat

\newcommand{\rrvert}{\vert}
\newcommand{\rrVert}{\Vert}
\newcommand{\llvert}{\vert}
\newcommand{\llVert}{\Vert}
\newcommand\Perp{\protect\mathpalette{\protect\independenT}{\perp}}
\def\independenT#1#2{\mathrel{\rlap{$#1#2$}\mkern4.1mu{#1#2}}}
\def\fraca#1#2{{(#1)}/{(#2)}}
\def\fracz#1#2{{#1}/{(#2)}}
\makeatother

\begin{document}
\begin{frontmatter}

\title{Causal Etiology of the Research of~James~M.~Robins}%
\runtitle{Robins' causal etiology}

\begin{aug}
\author[A]{\fnms{Thomas S.} \snm{Richardson}\corref{}\ead[label=e1]{thomasr@uw.edu}}
\and
\author[B]{\fnms{Andrea} \snm{Rotnitzky}\ead[label=e2]{arotnitzky@utdt.edu}}
\runauthor{T. S. Richardson and A. Rotnitzky}

\affiliation{University of Washington and Universidad Torcuato Di Tella
\& CONICET}

\address[A]{Thomas S. Richardson is Professor and Chair, Department of Statistics, University of Washington,
Box 354322, Seattle, Washington 98195, USA \printead{e1}.}
\address[B]{Andrea Rotnitzky is Professor, Department of Economics, Universidad Torcuato Di Tella \&
CONICET, Av. Figueroa Alcorta 7350,
S\'{a}enz Valiente 1010, Buenos Aires, Argentina \printead{e2}.}
\end{aug}

\begin{abstract}
This issue of \textit{Statistical Science} draws its inspiration from the work of
James M.~Robins. Jon Wellner, the Editor at the time, asked the two of
us to
edit a special issue that would highlight the research topics studied by
Robins and the breadth and depth of Robins' contributions. Between
the two of us, we have collaborated closely with Jamie for nearly 40 years.
We agreed to edit this issue because we recognized that we were among the
few in a position to relate the trajectory of his research career to
date.\renewcommand{\thefootnote}{\normalfont{1}}{\footnotemark[1]}%
\end{abstract}

\end{frontmatter}

\setcounter{footnote}{1}
\footnotetext{%
Here, we restrict attention to Robins' contributions to the research
literature. Robins has also contributed by training and mentoring leading
researchers in causal inference: among others, Elizabeth Halloran,
Miguel Hern\'{a}n, Eric
Tchetgen Tchetgen and Tyler VanderWeele worked with him as
graduate students. Both of the editors of this Special Issue were greatly
influenced by Robins' as a graduate student and post-doc and have been
fortunate to have subsequently collaborated with him over many years.}

Many readers may be unfamiliar with Robins' singular career trajectory and
in particular how his early practical experience motivated many of the
inferential problems with which he was subsequently involved. Robins majored
in mathematics at Harvard College, but then, in the spirit of the times,
left college to pursue more activist social and political goals. Several
years later, Robins enrolled in Medical School at Washington University in
St.~Louis, graduating in 1976. His M.D.~degree remains his only degree,
other than his high school diploma.

After graduating, he interned in medicine at Harlem Hospital in New York.
After completing the internship, Robins spent a year working as a primary
care physician in a community clinic in the Roxbury neighborhood of Boston.
During that year, he helped organize a vertical Service Employees
International Union affiliate that included all salaried personnel, from
maintenance to physicians, working at the health center. In
retaliation, he
was dismissed by the director of the clinic and found that he was somewhat
unwelcome at the other Boston community clinics. Unable to find a job and
with his unemployment insurance running out, he surprisingly was able to
obtain a prestigious residency in Internal Medicine at Yale University, a
testament, he says with some irony, to the enduring strength of one's Ivy
League connections.

At Yale, Robins and his college friend Mark Cullen, now head of General
Medicine at Stanford Medical School, founded an occupational health clinic,
with the goal of working with trade unions in promoting occupational health
and safety. When testifying in workers' compensation cases, Robins was
regularly asked whether it was ``more probable than not that
a worker's death or illness was \textit{caused} by exposure to
chemicals in
the workplace.'' Robins' lifelong interest in causal
inference began with his need to provide an answer. As the relevant
scientific papers consisted of epidemiologic studies and biostatistical
analyses, Robins enrolled in biostatistics and epidemiology classes at Yale.
He was dismayed to learn that the one question he needed to answer was the
one question excluded from formal discussion in the mainstream
biostatistical literature.\footnote{%
\citeauthor{Robi:Gree:esti:1989} (\citeyear{Robi:Gree:esti:1989,Robi:Gree:prob:1989})
provided a formal
definition of the probability of causation and a definitive answer to the
question in the following sense. They proved that the probability of
causation was not identified from epidemiologic data even in the
absence of
confounding, but that sharp upper and lower bounds could be obtained.
Specifically, under the assumption that a workplace exposure was never
beneficial, the probability $P(t)$ that a workers death occurring $t$ years
after exposure was due to that exposure was sharply upper bounded by
$1$ and
lower bounded by $\max [ 0,\{f_{1}(t)-f_{0}(t)\}/f_{1}(t) ] $,
where $f_{1}(t)$ and $f_{0}(t)$ are, respectively, the marginal
densities in
the exposed and unexposed cohorts of the random variable $T$ encoding time
to death.} At the time, most biostatisticians insisted that evidence for
causation could only be obtained through randomized controlled trials;
since, for ethical reasons, potentially harmful chemicals could not be
randomly assigned, it followed that statistics could play little role in
disentangling causation from spurious correlation.

\section{Confounding}

In his classes, Robins was struck by the gap present between the informal,
yet insightful, language of epidemiologists such as
\citeauthor{miettinen:1981} (\citeyear{miettinen:1981})
expressed in terms of ``confounding, comparability, and bias,'' and the
technical language of mathematical statistics in which these terms either
did not have analogs or had other meanings. Robins' first major paper ``The
foundations of confounding in Epidemiology'' written in 1982, though only
published in \citeyear{robins:foundation:1987}, was an attempt to bridge
this gap. As one example, he offered a precise mathematical definition for
the informal epidemiologic concept of a ``confounding variable'' that has
apparently stood the test of time
(see \citeauthor{vanderweele2013}, \citeyear{vanderweele2013}). As
a second example,
\citeauthor{efron:hinkley:78} (\citeyear{efron:hinkley:78}) had formally considered inference
accurate to order $n^{-3/2}$ in variance conditional on exact or approximate
ancillary statistics. Robins showed, surprisingly, that long before their
paper, epidemiologists had been intuitively and informally referring to an
estimator as ``unbiased'' just when it was asymptotically unbiased conditional
on either exact or approximate ancillary statistics; furthermore, they
intuitively required that the associated conditional Wald confidence
interval be accurate to $O(n^{-3/2})$ in variance. As a third example, he
solved the problem of constructing the tightest Wald-type intervals
guaranteed to have conservative coverage for the average causal effect among
the $n$ study subjects participating in a completely randomized experiment
with a binary response variable; he showed that this interval can be
strictly narrower than the usual binomial interval even under the Neyman
null hypothesis of no average causal effect. To do so, he constructed an
estimator of the variance of the empirical difference in treatment means
that improved on a variance estimator earlier proposed by %
\citeauthor{neyman:sur:1923} (\citeyear{neyman:sur:1923}).
\citeauthor{aronow2014} (\citeyear{aronow2014}) have recently generalized this
result in several directions including to nonbinary responses.

\section{Time-dependent Confounding and the \lowercase{g}-formula}\label{sec:time-dependent}

It was also in 1982 that Robins turned his
attention to the subject that would become his grail: causal inference from
complex longitudinal data with time-varying treatments,
that eventually culminated in his revolutionary papers %
\citeauthor{robins:1986} (\citeyear{robins:1986,robins:1987:addendum}).
His interest in this topic was
sparked by (i) a paper of
\citeauthor{gilbert:1982} (\citeyear{gilbert:1982})\footnote{%
The author, Ethel Gilbert, is the mother of Peter Gilbert who is a
contributor to this special issue; see
(\cite*{richardson:hudgens:2014}).} on the healthy worker survivor effect in
occupational epidemiology, wherein the author raised a number of questions
Robins answered in these papers and (ii) his medical experience of trying
to optimally adjust a patient's treatments in response to the evolution of
the patient's clinical and laboratory data.

\subsection{Overview}

Robins career from this point on became a ``quest'' to solve this
problem, and
thereby provide methods that would address central epidemiological
questions, for example, \textit{is a given long-term exposure harmful
or a
treatment beneficial?} \textit{If beneficial, what interventions, that is,
treatment strategies, are optimal or near optimal?}

In the process, Robins created a ``bestiary'' of causal models and analytic
methods.\footnote{In the epidemiologic literature, this bestiary is sometimes
referred to as the collection of ``g-methods.''} There are the basic ``phyla'' consisting of
the g-formula, marginal structural models and structural
nested models. These phyla then contain ``species,'' for example,
structural nested failure time models, structural nested distribution
models, structural nested (multiplicative
additive and logistic) mean models and yet further ``subspecies'':
direct-effect structural nested
models and optimal-regime structural nested models.

Each subsequent model in this taxa was developed to help answer
particular causal
questions in specific contexts that the ``older siblings'' were not quite
up to.
Thus, for example,
Robins' creation of structural nested and marginal structural models was
driven by the so-called null paradox, which could lead to
falsely finding a treatment effect where none existed, and was
a serious nonrobustness of the
estimated g-formula, his then current methodology.
Similarly, his research on
higher-order influence function estimators was motivated by a concern that,
in the presence of confounding by continuous, high dimensional confounders,
even doubly robust methods might fail to adequately control for
confounding bias.

This variety also reflects Robins' belief that the best analytic
approach varies with the causal question to be
answered, and, even more importantly, that confidence in one's substantive
findings only comes when multiple, nearly orthogonal, modeling strategies
lead to the same conclusion.

\subsection{Causally Interpreted Structured Tree Graphs}
\label{sec:tree-graph}

Suppose one wishes to estimate from longitudinal data the causal effect of
time-varying treatment or exposure, say cigarette smoking, on a failure time
outcome such as all-cause mortality. In this setting, a~time-dependent
confounder is a time-varying covariate (e.g.,~presence of emphysema)
that is
a predictor of both future exposure and of failure. In 1982, the standard
analytic approach was to model the conditional probability (i.e., the hazard)
of failure time $t$ as a function of past exposure history using a
time-dependent Cox proportional hazards model. Robins formally showed that,
even when confounding by unmeasured factors and model specification are
absent, this approach may result in estimates of effect that may fail to
have a causal interpretation, regardless of whether or not one also adjusts
for the measured time-dependent confounders in the analysis. In fact, if
previous exposure also predicts the subsequent evolution of the
time-dependent confounders (e.g.,~since smoking is a cause of emphysema, it
predicts this disease) then the standard approach can find an artifactual
exposure effect even under the sharp null hypothesis of no net, direct or
indirect effect of exposure on the failure time of any subject.

Prior to \citeauthor{robins:1986} (\citeyear{robins:1986}),
although informal discussions of net, direct
and indirect (i.e.,~mediated) effects of time varying exposures were to be
found in the discussion sections of most epidemiologic papers, no formal
mathematical definitions existed.
To address this, \citeauthor{robins:1986} (\citeyear{robins:1986})
introduced a new counterfactual model,
the \textit{finest fully randomized causally interpreted structured tree graph}
(FFRCISTG)\footnote{A complete \hyperref[acronyms]{list} of acronyms
used is given
before the References.} model that extended the point treatment
counterfactual model
of %
\citeauthor{neyman:sur:1923} (\citeyear{neyman:sur:1923}) and
\citeauthor{rubin:estimating:1974} (\citeyear{rubin:estimating:1974,Rubi:baye:1978})\footnote{%
See \citeauthor{Freedman01122006} (\citeyear{Freedman01122006}) and
\citeauthor{sekhon2008neyman} (\citeyear{sekhon2008neyman}) for
historical reviews of the counterfactual point treatment model.}
to longitudinal studies with time-varying
treatments, direct and indirect effects and feedback of one cause on
another. Due to his lack of formal statistical
training, the notation and formalisms in
\citeauthor{robins:1986} (\citeyear{robins:1986}) differ from
those found in the mainstream literature; as a consequence the paper
can be
a difficult read.\footnote{%
Robins published an informal, accessible, summary of his main results
in the
epidemiologic literature (\cite*{robins:simpleversion:1987}). However, it was
not until \citeyear{robins:1992} (and many rejections) that his work on
causal inference with time-varying treatments appeared in a major
statistical journal.} 
\citeauthor{richardson:robins:2013} (\citeyear{richardson:robins:2013}, Appendix C)
present the FFRCISTG model using a more familiar
notation.\footnote{%
The perhaps more familiar \textit{Non-Parametric Structural Equation Model with
Independent Errors} (NPSEM-IE) considered by Pearl may be viewed as submodel
of Robins' FFRCISTG.

A \textit{Non-Parametric Structural Equation Model} (NPSEM)
assumes that all variables ($V$) can be intervened
on. In contrast, the FFRCISTG model does not require one to assume this.
However, if all variables in $V$ can be intervened on, then the FFRCISTG
specifies a set of one-step ahead counterfactuals,
$V_{m}(\overline{v}_{m-1}) $ which may equivalently be written as structural equations
$V_{m}(\overline{v}_{m-1})=f_{m}(\overline{v}_{m-1},\varepsilon_{m})$ for
functions $f_{m}$ and (vector-valued) random errors $\varepsilon_{m}$. Thus, leaving
aside notational differences, structural equations and one-step ahead
counterfactuals are equivalent. All other counterfactuals, as well as
factual variables, are then obtained by recursive substitution.

However, the NPSEM-IE model of Pearl (\citeyear{pearl:2000}) further assumes the errors
$\varepsilon_{m}$ are jointly independent. In contrast, though an FFRCISTG model is also
an NPSEM, the errors (associated with incompatible counterfactual
worlds) may
be dependent---though any such dependence could not be detected in a RCT.
Hence, Pearl's model is a strict submodel of an FFRCISTG model.}


\begin{figure}[b]

\includegraphics{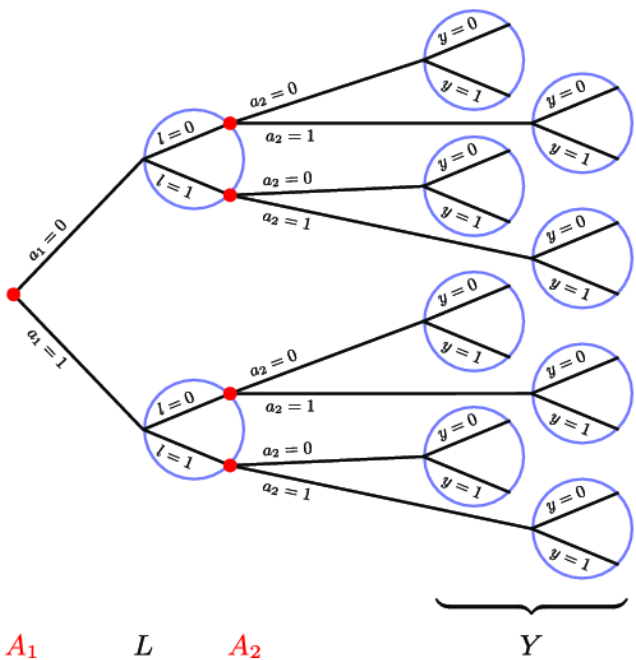}

\caption{Causal tree graph depicting a simple scenario with treatments at two times
$A_1$, $A_2$, a response $L$ measured prior to $A_2$, and a final
response $Y$. Blue circles indicate evolution of the process determined by Nature;
red dots indicate potential  treatment choices.}
\label{fig:event-tree}
\end{figure}

We illustrate the basic ideas using a simplified example. Suppose that we
obtain data from an observational or randomized study in which $n$
patients are treated at two times. Let $A_{1}$ and $A_{2}$ denote the
treatments. Let $L$ be a measurement taken just prior to the second
treatment and let $Y$ be a final outcome, higher values of which are
desirable. To simplify matters, for now we will suppose that all of the
treatments and responses are binary. As a concrete example, consider a study
of HIV infected subjects with $(A_{1},L,A_{2},Y)$, respectively, being
binary indicators of anti-retroviral treatment at time~$1$, high CD4
count just
before time $2$, anti-retroviral therapy at time $2$, and survival at
time $3$
(where for simplicity we assume no deaths prior to assignment of $A_2$).
There are $2^{4}=16$ possible observed data sequences for
$(A_{1},L,A_{2},Y)$; these may be depicted as an event tree as in
Figure~\ref{fig:event-tree}.\hskip.2pt%
\footnote{%
In practice, there will almost always exist baseline covariates measured
prior to $A_{1}$. In that case, the analysis in the text is to be understood
as being with a given joint stratum of a set of baseline covariates
sufficient to adjust for confounding due to baseline factors.} 
\citeauthor{robins:1986} (\citeyear{robins:1986})
referred to such event trees as ``structured tree graphs.''

We wish to assess the effect of the two treatments $(a_1, a_2)$ on $Y$.
In more detail,
for a given subject we suppose the existence of four potential \mbox{outcomes}
$Y(a_{1},a_{2})$ for $a_{1},a_{2}\in\{0,1\}$ which are the outcome a patient
would have if (possibly counter-to-fact) they were to receive the treatments
$a_{1}$ and $a_{2}$. Then $E[Y(a_{1},a_{2})]$ is the mean outcome (e.g.,~the
survival probability) if everyone in the population were to receive the
specified level of the two treatments. The particular instance of this
regime under which everyone is treated at both times, so
$a_{1}=a_{2}=1$, is
depicted in Figure~\ref{fig:event-tree-reg}(a). We are interested in
estimation of these four means since the regime $(a_{1},a_{2})$ that
maximizes $E[Y(a_{1},a_{2})]$ is the regime a new patient exchangeable
with the $n$ study subjects should follow.

There are two extreme scenarios: If in an observational study, the
treatments are assigned, for example, by doctors, based on additional unmeasured
predictors $U$ of $Y$ then $E[Y(a_{1},a_{2})]$ is not identified since those
receiving $(a_{1},a_{2})$ within the study are not representative of the
population as a whole.

At the other extreme, if the data comes from a completely randomized
clinical trial (RCT) in which treatment is assigned independently at each
time by the flip of coin, then it is simple to see that the
counterfactual $Y( a_{1},a_{2}) $ is independent of the treatments
$ (A_{1},A_{2} ) $ and that the average potential outcomes are identified
since those receiving $(a_{1},a_{2})$ in the study are a simple random
sample of the whole population. Thus,
%
\begin{eqnarray}
Y( a_{1},a_{2}) &\Perp & \{ A_{1},A_{2}
\} , \label{eq:full-rand}
\\
E\bigl[Y(a_{1},a_{2})\bigr] &=& E[ Y\mid A_{1}
= a_{1},A_{2} = a_{2}], \label{eq:asscaus}
\end{eqnarray}
where the right-hand side of (\ref{eq:asscaus}) is a function of the
observed data distribution. In a
completely randomized experiment, association is causation: the
associational quantity on the right-hand side of (\ref{eq:asscaus})
equals the causal quantity on the left-hand side.

Robins, however, considered an intermediate trial design in which both
treatments are randomized, but the probability of receiving $A_{2}$ is
dependent on both the treatment received initially ($A_{1}$) and the
observed response ($L$); a scenario now termed a \textit{sequential
randomized trial}. Robins viewed his analysis as also applicable to
observational data as follows. In an observational study, the role of an
epidemiologist is to use subject matter knowledge to try to collect in $L$
sufficient data to eliminate confounding by unmeasured factors, and
thus to
have the study mimic a sequential RCT. If successful, the only difference
between an actual sequential randomized trial and an observational
study is
that in the former the randomization probabilities
$\Pr(A_{2}=1 \mid  L,A_{1})$
are known by design while in the latter they must be estimated from the
data.\footnote{%
Of course, one can never be certain that the epidemiologists were successful
which is the reason RCTs are generally considered the gold standard for
establishing causal effects.} 

Robins viewed the sequential randomized trial as a collection of five
trials in total: the original
trial at $t=1$, plus a set of four randomized trials at $t=2$ nested
within the original trial.\footnote{That is, the trials
starting at $t=2$ are on study populations defined by specific
$(A_1,L)$-histories.}
Let the counterfactual $L( a_{1}) $ be the outcome $L$
when $A_{1}$ is set to $a_{1}$. Since the counterfactuals $Y(a_{1},a_{2})$
and $L( a_{1}) $ do not depend on the actual treatment received,
they can be viewed, like a subject's genetic make-up, as a fixed (possibly
unobserved) characteristic of a subject and therefore independent of the
randomly assigned treatment conditional on pre-randomization covariates.
That is, for each $(a_{1},a_{2})$ and $l$:
%
\begin{eqnarray}
\bigl\{ Y(a_{1},a_{2}),L(a_{1}) \bigr\} & \Perp
& A_{1}, \label{eq:ind1}
\\
Y(a_{1},a_{2})& \Perp & A_{2} \mid
A_{1} = a_{1},\quad L = l. \label{eq:ind2}
\end{eqnarray}

These independences suffice to identify the joint density
$f_{Y(a_{1},a_{2}),L(a_{1})}(y,l)$ of $(Y(a_{1},a_{2}), L(a_{1}))$
from the
distribution of the factual variables by the ``g-computation algorithm
formula'' (or simply \textit{g-formula}) density
%
\begin{equation}
f_{a_{1},a_{2}}^{\ast}(y,l)\equiv f(y \mid  a_{1},l,a_{2})f(l
\mid  a_{1})
\end{equation}
provided the conditional probabilities on the right-hand side are well-defined
(\citeauthor{robins:1986}, \citeyear{robins:1986}, page 1423).
Note that $f_{a_{1},a_{2}}^{\ast}(y,l)$ is obtained from the joint density of the
factuals by removing the treatment terms
$f(a_{2} \mid a_{1},l,a_{2})f(a_{1})$. This is in-line with the
intuition that $A_{1}$ and $A_{2}$ cease to be
random since, under the regime, they are set by intervention to
constants $a_{1}$ and $a_{2}$. The g-formula was later referred to as the ``manipulated
density'' by \citeauthor{cps93} (\citeyear{cps93})
and the ``truncated factorization'' by
\citeauthor{pearl:2000} (\citeyear{pearl:2000}).

\citeauthor{robins:1987:addendum} (\citeyear{robins:1987:addendum})
showed that under the weaker condition that
replaces (\ref{eq:ind1}) and (\ref{eq:ind2}) with
%
\begin{eqnarray}
\label{eq:statrand} Y(a_{1},a_{2}) &\Perp & A_{1}
\quad\hbox{and}
\nonumber
\\[-8pt]
\\[-8pt]
Y(a_{1},a_{2}) &\Perp& A_{2} \mid  A_{1}
= a_{1}, \quad L = l,
\nonumber
\end{eqnarray}
the marginal density of $Y(a_{1},a_{2})$ is still identified by
%
\begin{equation}
\label{eq:g-formula-for-y} f_{a_{1},a_{2}}^{\ast}(y)=\sum
_{l}f(y \mid  a_{1},l,a_{2})f(l \mid
a_{1}),
\end{equation}
the marginal under $f_{a_{1},a_{2}}^{\ast}(y,l)$.\footnote{%
The g-formula density for $Y$ is a generalization of standardization of
effect measures to time varying treatments. See
\citeauthor{keiding:2014} (\citeyear{keiding:2014})
for a historical review of standardization.} Robins called (\ref{eq:statrand})
\textit{randomization w.r.t.~$Y$}.\footnote{%
Note that the distribution of $L(a_{1})$ is no longer identified under this
weaker assumption.}
Furthermore, he provided substantive examples of
observational studies in which only the weaker assumption would be expected
to hold. It is much easier to describe these studies using representations
of causal systems using Directed Acyclic Graphs and Single World
Intervention Graphs, neither of which existed when
(\cite*{robins:1987:addendum}) was written.

\begin{figure*}

\includegraphics{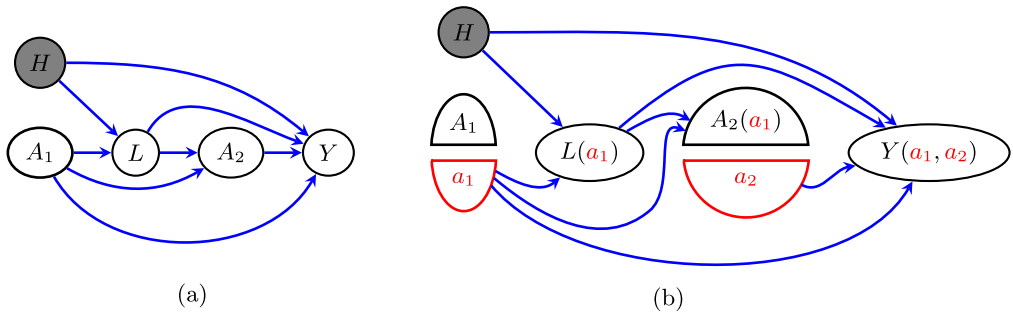}

\caption{\textup{(a)} A causal DAG $\mathcal{G}$ describing a sequentially
randomized trial; \textup{(b)} the SWIG $\mathcal{G}(a_1,a_2)$
resulting from intervening on $A_1$
and $A_2$.}\label{fig:seq-rand}
\end{figure*}

\subsection{Causal DAGs and Single World Intervention Graphs (SWIGs)}
\label{sec:dags}

Causal DAGs were first introduced in the seminal work of
\citeauthor{cps93} (\citeyear{cps93}); the
theory was subsequently developed and extended by
\citeauthor{pearl:biom} (\citeyear{pearl:biom,pearl:2000}) among others.

A causal DAG with random variables $V_{1},\ldots,V_{M}$ as nodes is a graph
in which (1) the lack of an arrow from node $V_{j}$ to $V_{m}$ can be
interpreted as the absence of a direct causal effect of $V_{j}$ on $V_{m}$
(relative to the other variables on the graph), (2) all common causes,
even if unmeasured, of any pair of variables on the graph are
themselves on
the graph, and (3) the Causal Markov Assumption (CMA) holds. The CMA links
the causal structure represented by the Directed Acyclic Graph (DAG) to
the statistical data obtained
in a study. It states that the distribution of the factual variables factor
according to the DAG. A distribution factors according to the DAG if
nondescendants of a given variable $V_{j}$ are independent of $V_{j}$
conditional on $\hbox{pa}_{j}$, the parents of $V_{j}$. The CMA is
mathematically equivalent
to the statement that the density $f(v_{1},\ldots,v_{M})$ of the variables
on the causal DAG $\mathcal{G}$ satisfies the Markov factorization
%
\begin{equation}
\label{eq:dag-factor} f(v_{1},\ldots,v_{M})=\prod
_{j=1}^{M}f(v_{j}\mid\mathrm{pa}_{j}).
\end{equation}
A graphical criterion, called d-separation
(\cite*{pearl:1988}),
characterizes all the marginal and conditional independences that hold in
every distribution obeying the Markov factorization (\ref{eq:dag-factor}).

Causal DAGs may also be used to represent the joint distribution of the
observed data under the counterfactual FFRCISTG model of
\citeauthor{robins:1986} (\citeyear{robins:1986}).
This follows because an FFRCISTG model over the
variables $\{V_{1},\ldots,V_{M}\}$ induces a
distribution that factors as (\ref{eq:dag-factor}).
Figure~\ref{fig:seq-rand}(a) shows a causal Directed Acyclic Graph (DAG)
corresponding to the sequentially randomized experiment described above:
vertex $H$ represents an unmeasured common cause (e.g.,~immune
function) of CD4
count $L$ and survival $Y$. Randomization of treatment implies $A_{1}$ has
no parents and $A_{2}$ has only the observed variables $A_{1}$ and $L$ as
parents.

\begin{figure*}[t]

\includegraphics{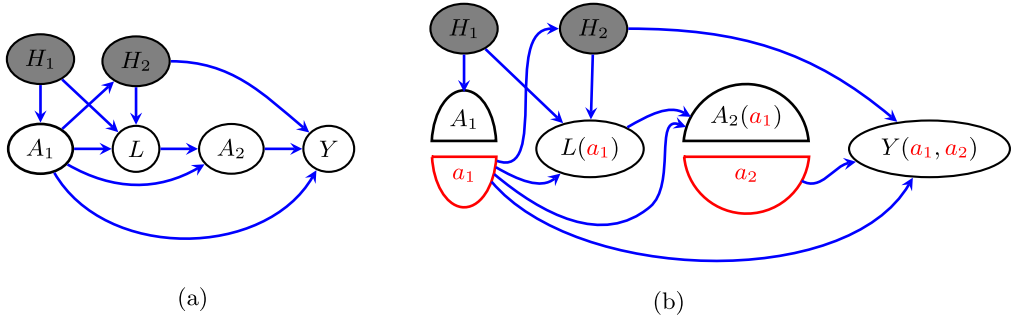}

\caption{Formaldehyde study: $H_{1}$, indicator of sensitivity to eye
irritants; $A_{1}$, formaldehyde exposure at time $1$; $H_{2}$, lung cancer;
$L$, current employment; $A_{2}$, formaldehyde exposure at time $2$;
$Y$, survival. $H_1$ and $H_2$ are unmeasured. \textup{(a)} A causal DAG
$\mathcal{G}$ in which initial treatment is
confounded, while the second treatment is sequentially randomized; \textup{(b)} the
SWIG $\mathcal{G}(a_1,a_2)$. $L$ is known to have no direct effect on $Y$
except indirectly via the effect on $A_2$; $H_1$ influences $A_1$ but
not $A_2$. See text for further explanation.}
\label{fig:seq-rand-variant}
\end{figure*}

Single-World Intervention Graphs (SWIGs), introduced in %
(\cite*{richardson:robins:2013}), provide a simple way to derive the
counterfactual independence relations implied by an FFRCISTG model. SWIGs
were designed to unify the graphical and potential outcome approaches to
causality. The nodes on a SWIG are the counterfactual random variables
associated with a specific hypothetical intervention on the treatment
variables. The SWIG in Figure~\ref{fig:seq-rand}(b) is derived from the
causal DAG in Figure~\ref{fig:seq-rand}(a) corresponding to a sequentially
randomized experiment. The SWIG represents the counterfactual world in which
$A_{1}$ and $A_{2}$ have been set to $(a_{1},a_{2})$, respectively.
\citeauthor{richardson:robins:2013} (\citeyear{richardson:robins:2013})
show that under the (naturally associated)
FFRCISTG model the distribution of the counterfactual variables on the SWIG
factors according to the graph. Applying Pearl's d-separation criterion to
the SWIG we obtain the independences (\ref{eq:ind1}) and
(\ref{eq:ind2}).%
\footnote{%
More precisely, we obtain the SWIG independence
$Y(a_{1},a_{2})\Perp A_{2}(a_{1}) \mid  A_{1},L(a_1)$, that implies (\ref{eq:ind2})
by the consistency assumption after instantiating $A_{1}$ at $a_{1}$. Note
when checking d-separation on a SWIG all paths containing red ``fixed'' nonrandom vertices, such as $a_1$, are treated as always being blocked (regardless of the conditioning set).}


\citeauthor{robins:1987:addendum} (\citeyear{robins:1987:addendum})
in one of the aforementioned substantive
examples described an observational study of the effect of formaldehyde
exposure on the mortality of rubber workers which can represented by the
causal graph in Figure~\ref{fig:seq-rand-variant}(a). (This graph cannot
represent a sequential RCT because the treatment variable $A_{1}$ and the
response $L$ have an unmeasured common cause.) Follow-up begins at time of
hire; time $1$ on the graph. The vertices $H_{1}$, $A_{1}$,
$H_{2}$, $L_{2}$, $A_{2}$, $Y$ are
indicators of sensitivity to eye irritants, formaldehyde exposure at
time $1$, lung cancer, current employment, formaldehyde exposure at time $2$ and
survival. Data on eye-sensitivity and lung cancer were not collected.
Formaldehyde is a known eye-irritant. The presence of an arrow from $H_{1}$
to $A_{1}$ but not from $H_{1}$ to $A_{2}$ reflects the fact that subjects
who believe their eyes to be sensitive to formaldehyde are given the
discretion to choose a job without formaldehyde exposure at time of
hire but
not later. The arrow from $H_{1}$ to $L$ reflects the fact that eye
sensitivity causes some subjects to leave employment. The arrows from $H_{2}$
to $L_{2}$ and $Y$ reflects the fact that lung cancer causes both death and
loss of employment. The fact that $H_{1}$ and $H_{2}$ are independent
reflects the fact that eye sensitivity is unrelated to the risk of lung
cancer.

From the SWIG in Figure~\ref{fig:seq-rand-variant}(b), we can see that
(\ref{eq:statrand}) holds so we have randomization with respect to $Y$ but
$L( a_{1}) $ is not independent of $A_{1}$. It follows that the
g-formula $f_{a_{1},a_{2}}^{\ast}(y)$ equals the density of $Y(a_{1},a_{2})$
even though (i) the distribution of $L( a_{1}) $ is not
identified and (ii) neither of the individual terms $f(l \mid  a_{1})$ and
$f(y \mid  a_{1},l,a_{2})$ occurring in the g-formula has a causal
interpretation.\footnote{%
Above we have assumed the variables $A_{1}$, $L$, $A_{2}$, $Y$ occurring in the
g-formula are temporally ordered. Interestingly,
\citeauthor{robins:1986} (\citeyear{robins:1986}, Section~11)
showed identification by the g-formula can require a nontemporal ordering.
In his analysis of the Healthy Worker Survivor Effect,
data were available on temporally
ordered variables $(A_{1},L_{1},A_{2},L_{2},Y)$ where the $L_{t}$ are
indicators of survival until time year $t$, $A_{t}$ is the indicator of
exposure to a lung carcinogen and, there exists substantive background
knowledge that carcinogen exposure at $t$ cannot cause death within a year.
Under these assumptions, Robins proved that equation
(\ref{eq:statrand}) was false if one
respected temporal order and chose $L$ to be $L_{1}$, but was true if one
chose $L=L_{2}$. Thus, $E[Y(a_{1},a_{2})]$ was identified by the
g-formula $f_{a_{1},a_{2}}^{\ast}(y)$ only for $L=L_{2}$. See %
(\citeauthor{richardson:robins:2013}, \citeyear{richardson:robins:2013}, page~54)  for further  details.}

Subsequently, \citeauthor{tian02general} (\citeyear{tian02general})
developed a graphical algorithm for
nonparametric identification that is ``complete'' in the sense that if the
algorithm fails to derive an identifying formula, then the causal quantity
is not identified
(\citeauthor{shpitser06id}, \citeyear{shpitser06id};
\citeauthor{huang06do}, \citeyear{huang06do}).
This algorithm strictly
extends the set of causal identification results obtained by Robins for
static regimes.


\subsection{Dynamic Regimes}\label{sec:dynamic-regimes}

\begin{figure*}

\includegraphics{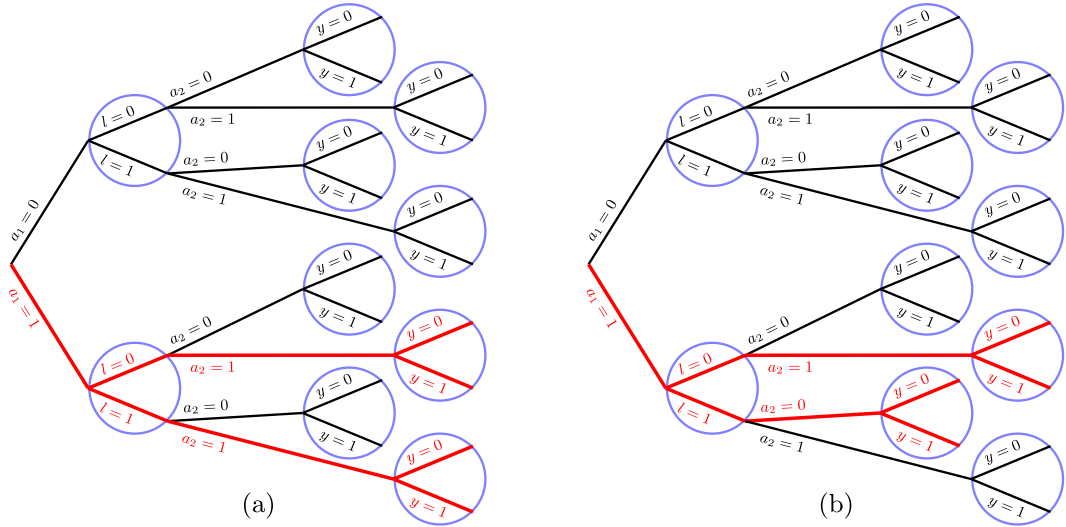}

\caption{Tree graphs depicting specific treatment regimes: \textup{(a)} $a_1=a_2=1$;
\textup{(b)} the dynamic regime $a_1=1$, $a_2=(1-l)$. The red paths indicate all
possible observed data sequences under these regimes.}
\label{fig:event-tree-reg}
\end{figure*}

The ``g'' in ``g-formula'' and elsewhere in Robins' work refers to generalized
treatment regimes $g$. The set $\mathbb{G}$ of all such regimes includes
\emph{dynamic} regimes in which a subject's treatment at time $2$
depends on
the response $L$ to the treatment at time $1$. An example of a dynamic
regime is the regime in which all subjects receive anti-retroviral
treatment at
time~$1$, but continue to receive treatment at time $2$ only if their CD4
count at time $2$ is low, indicating that they have not yet responded to
anti-retroviral treatment. In our study with no baseline covariates and $A_{1}$
and $A_{2}$ binary, a~dynamic regime $g$ can be written as
$g= (a_{1},g_{2}( l)  ) $ where the function $g_{2}(l)$ specifies
the treatment to be given at time $2$. The dynamic regime above has
$(a_{1} = 1,g_{2}(l) = 1-l)$ and is highlighted in
Figure~\ref{fig:event-tree-reg}. If $L$ is binary,
then $\mathbb{G}$ consists of $8$
regimes comprised of the $4$ earlier static regimes
$ (a_{1},a_{2} ) $ and $4$ \textit{dynamic} regimes. The \textit{g-formula}
density associated with a regime $g= ( a_{1},g_{2}(l) ) $ is
\[
f_{g}^{\ast}(y,l)\equiv f( l \mid  a_{1})f\bigl(y \mid
A_{1}=a_{1},L=l,A_{2}=g_{2}( l)
\bigr).
\]
Letting $Y(g)$ be a subject's counterfactual outcome under regime $g$, %
\citeauthor{robins:1987:addendum} (\citeyear{robins:1987:addendum}) proves that if both of the following hold:
%
\begin{eqnarray}
\label{eq:indg} Y(g)& \Perp & A_{1},
\nonumber
\\[-8pt]
\\[-8pt]
Y(g)& \Perp & A_{2} \mid  A_{1} = a_{1}, \quad L =
l
\nonumber
\end{eqnarray}
then $f_{Y(g)}(y)$ is identified by the g-formula density for $Y$:
\begin{eqnarray*}
f_{g}^{\ast}(y) &=&\sum_{l}f_{g}^{\ast}(y,l)
\\
&=&\sum_{l}f\bigl(y \mid  A_{1}=a_{1},L=l,A_{2}=g_{2}(l) \bigr)
\\\
&&\hspace*{15pt}{} \cdot f(l \mid  a_{1}).
\end{eqnarray*}
\citeauthor{robins:1987:addendum} (\citeyear{robins:1987:addendum})
refers to (\ref{eq:indg}) as the assumption
that regime $g$ \textit{is randomized with respect to $Y$}. Given a
causal DAG, Dynamic SWIGs (dSWIGS) can be used to check whether
(\ref{eq:indg}) holds.
\citeauthor{tian:dynamic:2008} (\citeyear{tian:dynamic:2008}) gives a complete graphical
algorithm for identification
of the effect of dynamic regimes based on DAGs.

Independences (\ref{eq:ind1}) and (\ref{eq:ind2}) imply that (\ref{eq:indg})
is true for all $g\in\mathbb{G}$. For a drug treatment, for which, say,
higher outcome values are better, the optimal regime
$g_{\mathrm{opt}}$ maximizing $E[ Y(g)] $ over $g\in\mathbb{G}$ is almost
always a dynamic regime, as treatment must be discontinued when
toxicity, a component of $L$, develops.

\citeauthor{robins:iv:1989} (\citeyear{robins:iv:1989,robins:1986}, page 1423)
used the g-nota\-tion $f(y \mid  g)$
as a shorthand for $f_{Y(g)}(y)$ in order to emphasize that this was the
density of $Y$ \textit{had intervention $g$ been applied to the population}.
In the special case of static regimes $ ( a_{1},a_{2} ) $, he wrote
$f(y \mid  g=(a_{1},a_{2}))$.\footnote{\citeauthor{pearl:biom} (\citeyear{pearl:biom})
introduced an
identical notation except that he substituted the word ``do'' for ``$g=$,'' thus
writing $f(y \mid  \operatorname{do}(a_{1},a_{2}))$.}

\subsection{Statistical Limitations of the Estimated g-Formulae}

Consider a sequentially randomized experiment. In this context, randomization
probabilities $f(a_{1})$ and $f(a_{2} \mid  a_{1},l)$ are known by design;
however, the densities $f(y \mid  a_{1},a_{2},l)$ and $f(l \mid  a_{1})$ are
not known and, therefore, they must be replaced by estimates
$\widehat{f}(y \mid  a_{1},a_{2},l_{2})$ and $\widehat{f}(l \mid  a_{1})$ in the g-formula.
If the sample size is moderate and $l$ is high dimensional, these
estimates must
come from fitting dimension-reducing models. Model misspecification
will then lead to
biased estimators of the mean of $Y(a_{1},a_{2})$.
\citeauthor{robins:1986} (\citeyear{robins:1986})
and \citeauthor{robins97estimation} (\citeyear{robins97estimation})
described a serious nonrobustness of the
g-formula: the so-called ``null paradox'': In biomedical trials, it is
frequently of interest to consider the possibility that the sharp causal
null hypothesis of no effect of either $A_{1}$ or $A_{2}$ on $Y$ holds.
Under this null, the causal DAG generating the data is as in
Figure \ref{fig:seq-rand} except without the arrows from $A_{1}$, $A_{2}$ and $L$ into
$Y$.\footnote{%
If the $L\rightarrow Y$ edge is present, then $A_{1}$ still has an
effect on $Y$.}
Then, under this null, although
$f_{a_{1},a_{2}}^{\ast}(y)=\sum_{l}f(y \mid  a_{1},l,a_{2})f(l \mid  a_{1})$
does not depend on $ ( a_{1},a_{2} )$, nonetheless
both $f(y \mid  a_{1},l,a_{2})$ and
$f(l \mid  a_{1})$ will, in general, depend on $a_{1}$ (as may be seen via
d-connection).%
\footnote{%
The dependence of $f(y \mid  a_{1},l,a_{2})$ on $a_{1}$ does not represent
causation but rather selection bias due to conditioning on the common effect
$L$ of $H_{1}$ and $A_{1}$.} In general, if $L$ has discrete
components, it
is not possible for standard nonsaturated parametric models (e.g., logistic
regression models) for both $f(y\mid a_{1},a_{2},l_{2})$ and $f(l_{2} \mid  a_{1})$
to be correctly specified, and thus depend on $a_{1}$ and yet for
$f_{a_{1},a_{2}}^{\ast}(y)$ not to depend on $a_{1}$.\footnote{But see %
\citeauthor{cox:wermuth:1999} (\citeyear{cox:wermuth:1999}) for another approach.}
As a consequence, inference based on the
estimated g-formula must result in the sharp null hypothesis being falsely
rejected with probability going to $1$, as the trial size increases,
even when it is true.

\subsection{\texorpdfstring{Structural Nested Models\protect\footnote{These models are discussed by
\citeauthor{joffe:vansteelandt:2014} (\citeyear{joffe:vansteelandt:2014}) in this
issue.}}{Structural Nested Models}}\label{sec:snm}

To overcome the null paradox,
\citeauthor{robins:iv:1989} (\citeyear{robins:iv:1989}) and %
\citeauthor{robins:pneumocystis:carinii:1992} (\citeyear{robins:pneumocystis:carinii:1992})
introduced the semiparametric
structural nested distribution model (SNDMs) for continuous outcomes
$Y$ and
structural nested failure time models (SNFTMs) for time to event outcomes.
See \citeauthor{robins:snftm} (\citeyear{robins:longdata,robins:snftm})
for additional details.

%
\citeauthor{robins:1986} (\citeyear{robins:1986}, Section~6)
defined the \emph{$g$-null hypothesis}
as
%
\begin{eqnarray}
\label{g-null} && H_{0}\dvtx \mbox{the distribution of }Y(g)
\nonumber
\\[-8pt]
\\[-8pt]
&& \hspace*{20pt} \mbox{is the same for all }g\in\mathbb{G}.
\nonumber
\end{eqnarray}
This hypothesis is implied by the sharp null hypothesis of no effect of
$A_{1}$ or $A_{2}$ on any subject's $Y$. If (\ref{eq:indg})
holds for all $g\in\mathbb{G,}$ then the $g$-null hypothesis is
equivalent to
any one of the following assertions:
\begin{enumerate}[(iii)]
\item[(i)] $f_{g}^{\ast}(y)$ equals the factual density $f( y) $\vspace*{2pt}
for all $g\in\mathbb{G}$;
\item[(ii)] $Y\Perp  A_1$ and $Y\Perp A_2 \mid  L,A_1$;
%

\item[(iii)] $f_{a_{1},a_{2}}^{\ast}(y)$ does not depend on
$ (a_{1},a_{2} ) $ and $Y\Perp A_{2} \mid  L,A_1$;
\end{enumerate}
see \citeauthor{robins:1986} (\citeyear{robins:1986}, Section~6).
In addition, any one of these assertions
exhausts all restrictions on the observed data distribution implied by the
sharp null hypothesis.

Robins' goal was to construct a causal model indexed by a
parameter $\psi^{\ast}$ such that in a sequentially randomized trial
(i) $\psi^{\ast}=0$ if and only if the $g$-null hypothesis (\ref{g-null})
was true and (ii) if known, one
could use the randomization probabilities to both construct an
unbiased estimating function for $\psi^{\ast}$ and to construct tests
of $\psi^{\ast}=0$ that were guaranteed (asymptotically) to reject under the
null at the nominal level. The SNDMs and SNFTMs accomplish this goal
for continuous
and failure time outcomes $Y$. %
\citeauthor{robins:iv:1989} (\citeyear{robins:iv:1989}) and
\citeauthor{robins:cis:correcting:1994} (\citeyear{robins:cis:correcting:1994}) also
constructed additive and multiplicative structural nested mean models
(SNMMs) which satisfied the above properties except with the \mbox{$g$-null}
hypothesis replaced by the \emph{$g$-null mean hypothesis}:
%
\begin{equation}
\label{g-null-mean} H_{0}\dvtx E\bigl[Y(g)\bigr]=E[Y]\quad\mbox{for all }g
\in\mathbb{G}.
\end{equation}
As an example, we consider an additive structural nested mean model. Define
\begin{eqnarray*}
&& \gamma(a_{1},l,a_{2})
\\
&&\quad=  E\bigl[ Y(a_{1},a_{2})-Y(a_{1},0) \mid
L=l,   A_{1} = a_{1},
\\
&&\hspace*{168pt} A_{2} = a_{2}\bigr]
\end{eqnarray*}
and
\[
\gamma(a_{1})=E\bigl[ Y(a_{1},0)-Y(0,0) \mid  A_{1}
= a_{1}\bigr] .
\]
Note $\gamma(a_{1},l,a_{2})$ is the effect of the last blip of
treatment $a_{2}$ at time $2$ among subjects with observed history
$ (a_{1},l,a_{2} ) $, while $\gamma( a_{1})$ is the effect of
the last blip of treatment $a_{1}$ at time~$1$ among subjects with
history $a_{1}$. An additive SNMM specifies parametric models
$\gamma(a_{1},l,a_{2};\psi_{2})$ and $\gamma(a_{1};\psi_{1})$ for these blip
functions with $\gamma(a_{1};0)=\gamma(a_{1},l,a_{2};0)=0$. Under the
independence assumptions (\ref{eq:indg}),
$H_{2}(\psi_{2})\* d(L,A_1) \{ A_{2}-E[ A_{2} \mid  L,A_{1}]  \} $
and
$H_{1}( \psi)  \{ A_{1}-E[ A_{1}]  \} $
are unbiased estimating functions
for the true $\psi^{\ast}$, where
$H_{2}( \psi_{2}) =Y-\gamma(A_{1},L,A_{2};\psi_{2})$,
$H_{1}(\psi)=H_{2}( \psi_{2}) -\gamma(A_{1};\psi_{1})$,
and $d(L,A_1)$ is a user-supplied function of the same dimension as
$\psi_2$.
Under the $g$-null mean hypothesis (\ref{g-null-mean}),
the SNMM is guaranteed to be correctly specified with $\psi^{\ast}=0$.
Thus, these estimating functions when evaluated at $\psi^{\ast}=0$,
can be used in the construction of an asymptotically $\alpha$-level
test of the $g$-null mean
hypothesis when $f(a_{1})$ and $f(a_{2} \mid  a_{1},l)$ are known (or are
consistently estimated).\footnote{%
In the literature, semiparametric estimation of the parameters of a SNM
based on such estimating functions is referred to as \mbox{``g-estimation.''}}
When $L$ is a high-dimensional vector, the parametric blip models may well be
misspecified when $g$-null mean hypothesis is false. However, because
the functions $\gamma(a_{1},l,a_{2})$ and $\gamma(a_{1})$ are nonparametrically
identified under assumptions (\ref{eq:indg}), one can
construct consistent tests of the correctness of the blip models
$\gamma(a_{1},l,a_{2};\psi_{2})$ and $\gamma(a_{1};\psi_{1})$. Furthermore, one
can also estimate the blip functions using cross-validation %
(\cite*{robins:optimal:2004}) and/or flexible machine learning methods in lieu
of a prespecified parametric model
(\cite*{van2011targeted}). A recent
modification of a
multiplicative SNMM, the structural nested cumulative failure time model,
designed for censored time to event outcomes has computational advantages
compared to a SNFTM, because, in contrast to a SNFTM,
parameters are estimated using an unbiased estimating function that is
differentiable in the model
parameters; see \citeauthor{picciotto2012structural} (\citeyear{picciotto2012structural}).

\citeauthor{robins:optimal:2004} (\citeyear{robins:optimal:2004})
also introduced optimal-regime  SNNMs drawing on
the seminal work of \citeauthor{Murp:opti:2003} (\citeyear{Murp:opti:2003})
on semiparametric methods for the estimation of optimal treatment strategies.
Optimal-regime SNNM estimation, called A-learning in computer science, can
be viewed as a semiparametric implementation of dynamic programming %
(\citeauthor{bellman:1957}, \citeyear{bellman:1957}).\footnote{%
Interestingly, \citeauthor{robins:iv:1989} (\citeyear{robins:iv:1989}, page 127 and App.~1),
unaware of Bellman's work, reinvented the method of dynamic programming but
remarked that, due to the difficulty of the estimation problem,
it would only be of theoretical interest for
finding the optimal dynamic regimes from longitudinal epidemiological data.}
Optimal-regime SNMMs differ from standard SNMMs only in that $%
\gamma(a_{1})$ is redefined to be
\begin{eqnarray*}
\gamma(a_{1}) &=& E\bigl[Y\bigl(a_{1},g_{2,\mathrm{opt}}
\bigl(a_{1},L(a_{1})\bigr)\bigr)
\\
&&\hspace*{11pt}{} - Y\bigl(0,g_{2,\mathrm{opt}}\bigl(0,L(0)\bigr)\bigr)
\mid  A_{1} =a_{1}\bigr],
\end{eqnarray*}
where $g_{2,\mathrm{opt}}(a_{1},l)=\arg\max_{a_{2}}\gamma(a_{1},l,a_{2})$ is the
optimal treatment at time $2$ given past history $ ( a_{1},l ) $.
The overall optimal treatment strategy $g_{\mathrm{opt}}$ is then
$ (a_{1,\mathrm{opt}},g_{2,\mathrm{opt}}( a_{1},l)  ) $ where
$a_{1,\mathrm{opt}}=\arg\max_{a_{1}}\gamma(a_{1})$. More on the estimation of optimal treatment
regimes can be found in
\citeauthor{schulte:2014} (\citeyear{schulte:2014})
in this volume.

\subsection{Instrumental Variables and Bounds for the Average Treatment
Effect}

\citeauthor{robins:iv:1989} (\citeyear{robins:iv:1989,robins1993analytic})
also noted that structural nested
models can be used to estimate treatment effects when assumptions
(\ref{eq:indg}) do not hold but data are available on a time dependent
instrumental variable. As an example, patients sometimes fail to fill their
prescriptions and thus do not comply with their prescribed treatment. In
that case, we can take $A_{j}=( A_{j}^{p},A_{j}^{d}) $ for each time $j$,
where $A_{j}^{p}$ denotes the treatment \textit{prescribed} and $A_{j}^{d}$
denotes the \textit{dose} of treatment actually received at time~$j$.
Robins defined $A_{j}^{p}$ to be \textit{an instrumental variable} if
(\ref{eq:indg}) still holds
after replacing $A_{j}$ by $A_{j}^{p}$ and for all subjects
$Y( a_{1},a_{2})$ depends on $a_{j}=( a_{j}^{p},a_{j}^{d}) $ only through the actual
dose $a_{j}^{d}$. Robins noted that unlike the case of full compliance
(i.e., $A_{j}^{p}=A_{j}^{d}$ with probability $1)$ discussed earlier, the treatment
effect functions $\gamma$ are not nonparametrically identified.
Consequently, identification can only be achieved by correctly specifying
(sufficiently restrictive) parametric models for~$\gamma$.

If we are unwilling to rely on such parametric assumptions, then the
observed data distribution only implies bounds for the $\gamma$'s.
In particular, in the setting of a point treatment randomized trial with
noncompliance and the instrument $A_{1}^{p}$ being the assigned
treatment, \citeauthor{robins:iv:1989} (\citeyear{robins:iv:1989})
obtained bounds on the average causal effect
$E[Y(a_{d}=1)-Y(a_{d}=0)]$ of the received treatment $A_{d}$. To the best
of our knowledge, this paper was
the first to derive bounds for nonidentified causal effects defined through
potential outcomes.\footnote{%
See also \citeauthor{Robi:Gree:prob:1989} (\citeyear{Robi:Gree:esti:1989,Robi:Gree:prob:1989}).}
The study of such bounds has become an active area of research.
Other early papers include
\citeauthor{manski:1990} (\citeyear{manski:1990}) and
\citeauthor{balke:pearl:1994} (\citeyear{balke:pearl:1994}).\footnote{%
\citeauthor{balke:pearl:1994} (\citeyear{balke:pearl:1994}) showed that
Robins' bounds were not sharp in the
presence of ``defiers'' (i.e.,~subjects who would never take the treatment
assigned) and derived sharp bounds in that case.}
See
\citeauthor{richardson:hudgens:2014} (\citeyear{richardson:hudgens:2014})
in this volume for a survey of recent research on bounds.

\subsection{Limitations of Structural Nested Models}\label{sec:limits-of-snms}

\citeauthor{robins00marginal} (\citeyear{robins00marginal})
noted that there exist causal questions for which SNMs are not
altogether satisfactory. As an example, for $Y$ binary,
\citeauthor{robins00marginal} (\citeyear{robins00marginal}) proposed a
structural nested logistic model in order to ensure estimates of the
counterfactual mean of $Y$ were between zero and one. However, he noted that
knowledge of the randomization probabilities did not allow one to construct
unbiased estimating function for its parameter $\psi^{\ast}$. More
importantly, SNMs do not directly model the final object of public health
interest---the distribution or mean of the outcome $Y$ as function of the
regimes~$g$---as these distributions are generally functions not only
of the
parameters of the SNM but also of the conditional law of the time dependent
covariates $L$ given the past history. In addition, SNMs constitute
a rather large conceptual leap from standard associational regression models
familiar to most statisticians.
\citeauthor{Robi:marg:1997} (\citeyear{Robi:marg:1997,robins00marginal})
introduced a new class of causal models, marginal structural models, that
overcame these particular difficulties. Robins also pointed out that MSMs
have their own shortcomings, which we discuss below.
\citeauthor{robins00marginal} (\citeyear{robins00marginal})
concluded that the best causal model to use will
vary with the causal question of interest.

\subsection{Dependent Censoring and Inverse Probability Weighting}
\label{sec:censoring}

Marginal Structural Models grew out of Robins' work on censoring and
\textit{inverse
probability of censoring weighted} (IPCW) estimators. Robins
work on dependent censoring was motivated by the familiar clinical
observation that patients who did not return to the clinic and were thus
censored differed from other patients on important risk factors, for example
measures of cardio-pulmonary reserve. In the 1970s and
1980s, the analysis of right censored data was a major area of statistical
research, driven by the introduction of the proportional hazards model
(\cite*{cox:1972:jrssb}; \cite*{Kalb:Pren:stat:1980})
and by martingale methods for their analysis %
(\cite*{Aalen:counting:1978};
\cite*{andersen:borgan:gill:keiding:1992};
\cite*{Flem:Harr:coun:1991}).
This research, however, was focused on independent censoring. An important
insight in \citeauthor{robins:1986} (\citeyear{robins:1986})
was the recognition that by reframing the
problem of censoring as a causal inference problem as we will now explain,
it was possible to adjust for dependent censoring with the \mbox{g-formula}.

\citeauthor{Rubi:baye:1978} (\citeyear{Rubi:baye:1978})
had pointed out previously that
counterfactual causal inference could be viewed as a missing
data problem. \citeauthor{robins:1986} (\citeyear{robins:1986}, page 1491)
recognized that the converse was
indeed also true: a missing data problem could be viewed as a problem in
counterfactual causal inference.\footnote{%
A viewpoint recently explored by
\citeauthor{mohan:pearl:tian:2013} (\citeyear{mohan:pearl:tian:2013}).}
Robins conceptualized right censoring as just another time dependent
``treatment'' $A_{t}$ and one's inferential goal as the
estimation of the outcome $Y$ under the static regime~$g$ ``never
censored.'' Inference based on the g-formula was then licensed
provided that censoring was explainable in the sense that
(\ref{eq:statrand}) holds. This approach to
dependent censoring subsumed independent censoring
as the latter is a special case of the former.

Robins, however, recognized once again that inference based on the estimated
g-formula could be nonrobust. To overcome this difficulty, %
\citep{robins:rotnitzky:recovery:1992}
introduced IPCW tests and estimators
whose properties are easiest to explain in the context of a two-armed
RCT of
a single treatment ($A_{1}$). The standard Intention-to-Treat (ITT) analysis
for comparing the survival distributions in the two arms is a log-rank test.
However, data are often collected on covariates, both pre- and
post-randomization, that are predictive of the outcome as well as (possibly)
of censoring. An ITT analysis that tries to adjust for
dependent-censoring by
IPCW uses estimates of the arm-specific hazards of censoring as
functions of
past covariate history. The proposed IPCW tests have the following two
advantages compared to the log rank test. First, if censoring is dependent
but explainable by the covariates, the log-rank test is not
asymptotically valid. In contrast, IPCW tests asymptotically
reject at their nominal level
provided the arm-specific hazard estimators are consistent. Second,
when censoring is independent, although both the IPCW tests and the log-rank
test asymptotically reject at their nominal level, the IPCW tests, by
making use of
covariates, can be more powerful than the log-rank test even against
proportional-hazards alternatives. Even under independent censoring tests
based on the estimated g-formula are not guaranteed to be
asymptotically $\alpha$-level, and hence are not robust.

To illustrate, we consider here an RCT with $A_{1}$ being the randomization
indicator, $L$ a post-randomization covariate, $A_{2}$ the indicator of
censoring and $Y$ the indicator of survival. For simplicity, we assume that
any censoring occurs at time $2$ and that there are no failures prior to
time $2$. The IPCW estimator $\widehat{\beta}$ of the ITT effect
$\beta^{\ast}=E[ Y \mid  A=1] -E[ Y \mid  A=0] $ is defined as
the solution to
%
\begin{equation}
\mathbb{P}_{n}\bigl[ I ( A_{2}=0 ) U(\beta)/\widehat{
\Pr}(A_{2}=0 \mid  L,A_{1})\bigr] =0, \hspace*{-25pt}
\end{equation}
where $U(\beta)= ( Y-\beta A_{1} )  ( A_{1}-1/2 ) $,
throughout $\mathbb{P}_{n}$ denotes the empirical mean operator and
$\widehat{\Pr}( A_{2}=0 \mid  L,A_{1}) $ is an estimator of the
arm-specific conditional probability of being uncensored.
When first introduced in 1992, IPCW estimators, even when taking the
form of simple
Horvitz--Thompson estimators, were met with both surprise and
suspicion as they violated the then widely held belief that one should never
adjust for a post-randomization variable affected by treatment in a RCT.

\subsection{Marginal Structural Models}
\label{sec:msm}

\citeauthor{robins1993analytic} (\citeyear{robins1993analytic}, Remark~A1.3, pages 257--258)
noted that, for any
treatment regime $g$, if randomization w.r.t.~$Y$, that is,
(\ref{eq:indg}), holds, $\Pr\{Y(g)>y\}$ can be estimated by IPCW if one defines a
person's censoring time as the first time he/she fails to take the treatment
specified by the regime. In this setting, he referred to IPCW as
\textit{inverse
probability of treatment weighted} (IPTW).
In actual longitudinal data in which either (i)~treatment $A_{k}$
is measured at many times $k$ or (ii) the $A_{k}$ are discrete with many
levels or continuous, one often finds that few study subjects follow any
particular regime.
In response, \citeauthor{Robi:marg:1997} (\citeyear{Robi:marg:1997,robins00marginal}) introduced
MSMs. These models address the aforementioned
difficulty by borrowing information across regimes.
Additionally, MSMs represent another response to the $g$-null paradox
complementary to Structural Nested Models.

To illustrate, suppose that in our example of Section~\ref{sec:time-dependent},
$A_{1}$ and $A_{2}$ now have many levels.
An instance of an MSM for the counterfactual
means $E[ Y(a_{1},a_{2})]$ is a model that specifies that
\[
\Phi^{-1}\bigl\{E\bigl[Y(a_{1},a_{2})\bigr]\bigr
\}=\beta_{0}^{\ast} + \gamma\bigl(a_{1},a_{2};
\beta_{1}^{\ast}\bigr),
\]
where $\Phi^{-1}$ is a given link function such as the logit, log, or
identity link and $\gamma( a_{1},a_{2};\beta_{1}) $ is a known
function satisfying $\gamma( a_{1},a_{2};0) =0$. In this model,
$\beta_{1}=0$ encodes the \textit{static-regime mean null hypothesis} that
%
\begin{equation}\label{static null}
H_{0}\dvtx E\bigl[ Y(a_{1},a_{2})\bigr] \mbox{ is the same for all } (a_{1},a_{2} ) .
\hspace*{-15pt}
\end{equation}
\citeauthor{Robi:marg:1997} (\citeyear{Robi:marg:1997})
proposed IPTW estimators $( \widehat{\beta}_{0},\widehat{\beta}_{1}) $
of $ ( \beta_{0}^{\ast},\beta_{1}^{\ast} )$.
When the treatment probabilities are known, these estimators are
defined as the solution to
%
\begin{eqnarray}\label{eq:msm2}
\qquad && \mathbb{P}_{n}\bigl[ \vphantom{\hat{P}}Wv(A_{1},A_{2})
\bigl( Y-\Phi\bigl\{\beta_{0}+\gamma(A_1,A_2;
\beta_{1})\bigr\} \bigr) \bigr]
\nonumber\\[-8pt]
\\[-8pt]
&&\quad=0
\nonumber
\end{eqnarray}
for a user supplied vector function $v(A_{1},A_{2})$ of the dimension
of $ ( \beta_{0}^{\ast},\beta_{1}^{\ast} ) $ where
\[
W=1/ \bigl\{ f( A_{1}) f(A_{2} \mid  A_{1},L) \bigr
\}.
\]
Informally, the product $f(A_{1})f(A_{2} \mid  A_{1},L)$ is the ``probability
that a subject had the treatment history he did indeed have.''%
\footnote{
IPTW estimators and IPCW estimators are essentially equivalent. For
instance, in the censoring example of Section~\ref{sec:censoring}, on
the event $A_{2}=0$ of being uncensored, the IPCW denominator
$\widehat{pr}(A_{2}=0 \mid L,A_{1}) $ equals $f(A_{2} \mid  A_{1},L)$, the IPTW
denominator.}
When the treatment probabilities are unknown, they are
replaced by estimators.

Intuitively, the reason why the estimating function of (\ref{eq:msm2})
has mean
zero at $ ( \beta_{0}^{\ast},\beta_{1}^{\ast} ) $ is as
follows: Suppose the data had been generated from a sequentially randomized
trial represented by DAG in Figure~\ref{fig:seq-rand}. We may create a
pseudo-population by making $1/\{f(A_{1})f( A_{2}\mid A_{1},L)\}$ copies of each study
subject. It can be shown that in the resulting pseudo-population
$A_{2}\Perp  \{ L,A_{1} \}$, and thus is
represented by the DAG in Figure~\ref{fig:seq-rand}, except with both
arrows into
$A_{2}$ removed. In the pseudo-population, treatment is
completely randomized (i.e.,~there is no confounding by either measured or
unmeasured variables), and hence causation is association. Further, the
mean of
$Y(a_{1},a_{2})$ takes the same value in the pseudo-population as in
the actual population. Thus if, for example,
$\gamma(a_{1},a_{2};\beta_{1}) =\beta_{1,1}a_{1}+\beta_{1,2}a_{2}$ and
$\Phi^{-1} $ is the identity link, we can estimate
$ ( \beta_{0}^{\ast},\beta_{1}^{\ast} ) $
by OLS in the pseudo-population. However, OLS
in the pseudo-population is precisely weighted least squares in the actual
study population with weights $1/\{f(A_{1})f( A_{2}\mid A_{1},L)\}$.\footnote{%
More formally, recall that under (\ref{eq:statrand}),
$E[ Y(a_{1},a_{2}) ] =\Phi\{ \beta_{0}^{\ast}+\gamma(a_{1},a_{2};\beta_{1}^{\ast}) \} $
is equal to the g-formula $\int yf_{ a_{1},a_{2} }^{\ast}( y) \,dy$.
Now, given the joint density of the
data $f( A_{1},L,A_{2},Y) $, define
\[
\widetilde{f}( A_{1},L,A_{2},Y) =f( Y\mid
A_{1},L,A_{2}) \widetilde{f}_{2}(A_{2})
f( L\mid A_{1}) \widetilde{f}_{1}( A_{1}),
\]
where $\widetilde{f}_{1}( A_{1}) \widetilde{f}_{2}( A_{2}) $ are
user-supplied densities chosen so that $\widetilde{f}$ is absolutely
continuous with respect to $f$. Since the g-formula depends on the joint
density of the data only through $f( Y \mid  A_{1},L,A_{2}) $ and
$f(L \mid  A_{1})$, then it is identical under $\widetilde{f}$ and under $f$.
Furthermore, for each $a_{1}$, $a_{2}$ the g-formula under $\widetilde{f}$ is
just equal to $\widetilde{E}[ Y \mid  A_{1}=a_{1},A_{2}=a_{2}] $
since, under $\widetilde{f}$, $A_{2}$ is independent of
$ \{L,A_{1} \}$. Consequently, for any $q( A_{1},A_{2})$
\begin{eqnarray*}
\everymath{\displaystyle}
\begin{array}{rcl}
0&= & \widetilde{E} \bigl[ q( A_{1},A_{2}) \bigl( Y-\Phi
\bigl\{ \beta _{0}^{\ast}+\gamma \bigl( A_{1},A_{2};
\beta_{1}^{\ast} \bigr) \bigr\} \bigr) \bigr]
\\[6pt]
& =& E \bigl[ q( A_{1},A_{2}) \bigl\{ \widetilde{f}(
A_{1}) \widetilde{f}( A_{2}) / \bigl\{ f( A_{1})
f(A_{2} \mid  A_{1},L) \bigr\} \bigr\}
\\[6pt]
&& \hspace*{79pt} {} \cdot
\bigl( Y-\Phi\bigl\{ \beta_{0}^{\ast} + \gamma\bigl(
A_{1},A_{2};\beta_{1}^{\ast}\bigr)\bigr\}
\bigr) \bigr],
\end{array}
\end{eqnarray*}
where the second equality follows from the Radon--Nikodym theorem. The result
then follows by taking $q( A_{1},A_{2}) =v(A_{1},A_{2})/
\{\widetilde{f}( A_{1}) \widetilde{f}( A_{2})\}$.}

\citeauthor{robins00marginal} (\citeyear{robins00marginal}, Section~4.3)
also noted that the weights $W$ can be replaced by
the so-called stabilized weights $SW= \{ f( A_{1})
f(A_{2} \mid  A_{1}) \} / \{ f( A_{1}) f(A_{2}\mid A_{1},L) \}$, and
described settings where, for efficiency reasons, using $SW$ is preferable
to using $W$.

MSMs are not restricted to models for the dependence of the mean of
$Y(a_{1},a_{2})$ on $ ( a_{1},a_{2} ) $. Indeed, one can consider
MSMs for the dependence of any functional of the law of
$Y(a_{1},a_{2})$ on
$ ( a_{1},a_{2} )$, such as a quantile or the hazard function
if $Y$
is a time-to-event variable. If the study is fully randomized, that is,
(\ref{eq:full-rand})~holds, then an MSM model for a given functional of the
law of $Y( a_{1},a_{2}) $ is tantamount to an associational model
for the same functional of the law of $Y$ conditional on $A_{1}=a_{1}$
and $%
A_{2}=a_{2}$. Thus, under (\ref{eq:full-rand}), the MSM model can be
estimated using standard methods for estimating the corresponding
associational model. If the study is only sequentially randomized, that
is, (\ref{eq:statrand}) holds but (\ref{eq:full-rand}) does not, then the model can
still be estimated by the same standard methods but weighting each subject
by $W$ or $SW$.

\citeauthor{robins00marginal} (\citeyear{robins00marginal}) discussed disadvantages of
MSMs compared to SNMs. Here, we summarize some of the main drawbacks.
Suppose (\ref{eq:indg}) holds for all $g \in\mathbb{G}$.
If the $g$-null hypothesis (\ref{g-null}) is false but the static
regime null
hypothesis that the law of $Y( a_{1},a_{2}) $ is the same for all
$ ( a_{1},a_{2} ) $ is true, then by (iii) of Section~\ref{sec:snm},
$f( y \mid  A_{1}=a_{1},A_{2}=a_{2},L=l) $ will depend on $a_{2}$ for some
stratum $ ( a_{1}, l )$ thus implying a causal effect of $A_{2}$
in that stratum; estimation of an SNM model would, but estimation of an MSM
model would not, detect this effect. A second drawback is that
estimation of MSM models,
suffers from marked instability and finite-sample
bias in the presence of weights $W$ that are highly variable and skewed.
This is not generally an issue in SNM estimation.
A third limitation of MSMs is that when (\ref{eq:statrand})
fails but an instrumental variable is
available, one can still consistently estimate the parameters of a SNM
but not of an MSM.\footnote{%
Note that, as observed earlier, in this case identification is achieved
through parametric assumptions made by the SNM.}

An advantage of MSMs over SNMs that was not discussed in
Section~\ref{sec:limits-of-snms} is the following. MSMs can be constructed that are
indexed by easily interpretable parameters that quantify the overall
effects of a subset of all possible dynamic regimes
(\cite*{Miguel:Robins:dominique:2005};
\cite*{van:Pete:caus:2007};
\citeauthor{orellana2010dynamic} (\citeyear{orellana2010dynamic,orellana2010proof}).
As an example consider a longitudinal study of HIV infected
patients with baseline CD4 counts exceeding 600 in which we wish to
determine the
optimal CD4 count at which to begin anti-retroviral treatment. Let $g_{x}$
denote the dynamic regime that specifies treatment is to be initiated the
first time a subject's CD4 count falls below $x$,
$x\in \{1,2,\ldots, 600 \} $. Let
$Y(g_{x})$ be the associated counterfactual response and
suppose few study subjects follow any given regime. If we assume
$E[Y(g_{x})]$ varies smoothly with $x$, we can specify and fit (by IPTW) a dynamic
regime MSM model $E[Y(g_{x})]=\beta_{0}^{\ast}+\beta_{1}^{\ast T}h(x)$
where, say, $h(x)$ is a vector of appropriate spline functions.

\section{Direct Effects}

Robins' analysis of sequential regimes leads immediately to the
consideration of direct effects. Thus, perhaps not surprisingly, all
three of the
distinct direct effect concepts that are now an integral part of the causal
literature are all to be found in his early papers. Intuitively, all the
notions of
direct effect consider whether ``the outcome ($Y$) would have
been different had cause ($A_{1}$) been different, but the level of
($A_{2}$) remained unchanged.'' The notions differ regarding the
precise meaning of $A_2$ ``remained
unchanged.''

\begin{figure*}

\includegraphics{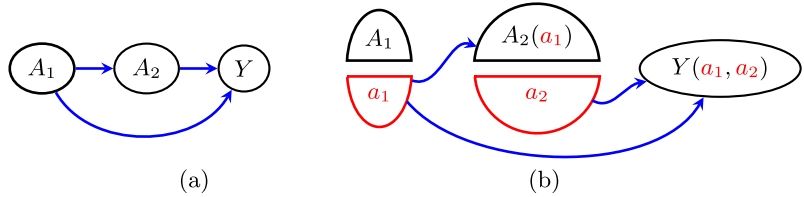}

\caption{\textup{(a)}  A causal DAG $\mathcal{G}$ with no (measured or unmeasured) confounding of $A_2$
 on $Y$; \textup{(b)} the SWIG $\mathcal{G}(a_1,a_2)$ resulting from intervening on $A_1$ and $A_2$.}\label{fig:no-confound}
\end{figure*}

\subsection{Controlled Direct Effects}\label{sec:cde}

In a setting in which there are temporally ordered treatments $A_{1}$
and $A_{2}$, it is natural to wonder whether the first treatment has any effect
on the final outcome were everyone to receive the second treatment.
Formally, we wish to compare the potential outcomes $Y(a_{1} = 1,a_{2} =1)$
and $Y(a_{1} = 0,a_{2} = 1)$.
\citeauthor{robins:1986} (\citeyear{robins:1986}, Section~8)
considered
such contrasts, that are now referred to as \textit{controlled direct
effects%
}. More generally, the \textit{average controlled direct effect of $A_{1}$
on $Y$ when $A_{2}$ is set to $a_{2}$} is defined to be
%
\begin{equation}\label{eq:acde}
\hspace*{23pt}
\mathrm{CDE}(a_{2})\equiv E\bigl[Y(a_{1}=1,a_{2})-Y(a_{1}=0,a_{2})\bigr] ,
\end{equation}
where $Y(a_{1}=1,a_{2})-Y(a_{1}=0,a_{2})$ is the individual level direct
effect. Thus, if $A_{2}$ takes $k$-levels then there are $k$ such contrasts.

Under the causal graph shown in Figure~\ref{fig:no-confound}(a), in
contrast to Figures~\ref{fig:seq-rand} and \ref{fig:seq-rand-variant}, the
effect of $A_{2}$ on $Y$ is unconfounded, by either measured or unmeasured
variables, association is causation and thus, under the associated FFRCISTG
model:
\begin{eqnarray*}
\mathrm{CDE}(a_{2}) &=& E[ Y \mid  A_{1}=1,A_{2}=a_{2}]
\\
&&{} - E[Y \mid A_{1}=0,A_{2}=a_{2}] .
\end{eqnarray*}

The CDE can be identified even in the presence of time-dependent
confounding. For example,
in the context of the FFRCISTG associated with either of the
causal DAGs shown in Figures~\ref{fig:seq-rand} and \ref%
{fig:seq-rand-variant}, the $\operatorname{CDE}(a_2)$ will be identified via the
difference in the
expectations of $Y$ under the g-formula densities $f_{a_1 =
1,a_2}^*(y)$ and $%
f_{a_1 = 0,a_2}^*(y)$.\footnote{%
See (\ref{eq:g-formula-for-y}).}

The CDE requires that the potential outcomes
$Y(a_{1},\allowbreak a_{2})$ be well-defined for all values of $a_{1}$ and $a_{2}$. This
is because the CDE treats both $A_{2}$ and $A_{1}$ as causes,
and interprets ``$A_{2}$ remained
unchanged'' to mean ``had there been an
intervention on $A_2$ fixing it to $a_2$.''

This clearly requires that the analyst be able to describe a well-defined
intervention on the mediating variable $A_{2}$.

There are many contexts in which there is no clear well-defined intervention
on $A_{2}$ and thus it is not meaningful to refer to $Y(a_{1},a_{2})$. The
CDE is not applicable in such contexts.

\subsection{Principal Stratum Direct Effects (PSDE)} \label{sec:psde}

\citeauthor{robins:1986} (\citeyear{robins:1986})
considered causal contrasts in the situation described
in Section~\ref{sec:censoring} in which death from a disease of
interest, for example,~a
heart attack, may be censored by death from other diseases. To describe these
contrasts, we suppose $A_{1}$ is a treatment of interest, $Y=1$ is the
indicator of death from the disease of interest (in a short interval
subsequent to a given fixed time $t$) and $A_{2}=0$ is the ``at risk
indicator'' denoting the absence of death either from other diseases or the
disease of interest prior to time $t$.

Earlier \citeauthor{Kalb:Pren:stat:1980} (\citeyear{Kalb:Pren:stat:1980})
had argued that if $A_{2}=1$, so that
the subject does not survive to time $t$, then the question of whether the
subject would have died of heart disease subsequent to $t$ had death
before $%
t$ been prevented is meaningless. In the language of counterfactuals, they
were saying (i) that if $A_{1}=a_{1}$ and $A_{2}\equiv A_{2}(a_{1}) =1$,
the counterfactual $Y(a_{1},a_{2}=0)$ is not well-defined
and (ii) the counterfactual $Y(a_{1},a_{2}=1)$ is never well-defined.

\citeauthor{robins:1986} (\citeyear{robins:1986}, Section~12.2)
observed that if one accepts this then the
only direct effect contrast that is well-defined is
$Y(a_{1} =1,a_{2}=0)- Y(a_{1} = 0,a_{2}=0)$ and that is well-defined only for those
subjects who would survive to $t$ regardless of whether they received
$a_{1} = 0$ or $a_{1} = 1$. In other words, even though $Y(a_{1},a_{2})$
may not be well-defined for all subjects and all $a_{1}$, $a_{2}$, the
contrast:
%
\begin{eqnarray}\label{eq:psde-contrast}
&& E\bigl[ Y(a_{1} = 0,a_{2})
 -Y(a_{1}= 1,a_{2}) \mid
\nonumber\\[-8pt]
\\[-8pt]
&&\hspace*{16pt}{}A_{2}(a_{1} =1)=A_{2}(a_{1} = 0)=a_2\bigr]
\nonumber
\end{eqnarray}
is still well-defined when $a_{2}=0$. As noted by Robins, this could provide
a solution to the problem of defining the causal effect of the
treatment $A_{1}$ on the outcome $Y$ in the context of censoring by death due to other
diseases.

\citeauthor{Rubi:more:1998} (\citeyear{Rubi:more:1998}) and
\citeauthor{Fran:Rubi:addr:1999} (\citeyear{Fran:Rubi:addr:1999,Fran:prin:2002}) later
used this same contrast to solve precisely the same problem of
``censoring by death.''\footnote{The analysis of %
\citeauthor{Rubi:dire:2004} (\citeyear{Rubi:dire:2004}) was also based on this contrast,
with $A_2$ no longer a failure time indicator so
that the contrast (\ref{eq:psde-contrast})
could be considered as well-defined for any value of $a_{2}$ for
which the conditioning event had positive probability.}

In the terminology of \citeauthor{Fran:prin:2002} (\citeyear{Fran:prin:2002})
for a subject
with $A_{2}(a_{1} = 1)=A_{2}(a_{1} = 0)=a_{2}$, the \textit{individual
principal stratum direct effect} is defined to be:\footnote{%
For subjects for whom $A_{2}(a_{1} = 1)\neq A_{2}(a_{1} = 0)$, no
principal stratum direct effect (PSDE) is defined.}
\[
Y(a_{1}=1,a_{2}) - Y(a_{1}=0,a_{2})
\]
(here, $A_{1}$ is assumed to be binary). The \textit{average PSDE in
principal stratum $a_{2}$} is then defined to be
%
\begin{eqnarray}\label{eq:psde2}
\qquad
\operatorname{PSDE}(a_{2})
&\equiv & E\bigl[Y(a_{1} = 1,a_{2}) -Y(a_{1} = 0,a_{2})\mid
\nonumber\\
&&\hspace*{16pt}
A_{2}(a_{1} = 1)=A_{2}(a_{1} = 0)=a_{2}\bigr]
\nonumber\\[-8pt]
\\[-8pt]
& =& E\bigl[ Y(a_{1} = 1)-Y(a_{1} = 0)\mid
\nonumber\\
&&\hspace*{13pt}
A_{2}(a_{1} = 1)=A_{2}(a_{1} = 0)=a_{2}\bigr],
\nonumber
\end{eqnarray}
where the second equality here follows, since
$Y(a_{1},\allowbreak A_{2}(a_{1}))=Y(a_{1})$.\footnote{This follows from
consistency.} In contrast to the CDE, the PSDE has the
advantage that it may be defined, via (\ref{eq:psde2}), without
reference to
potential outcomes involving intervention on $a_{2}$. Whereas the CDE
views $%
A_{2}$ as a treatment, the PSDE treats $A_{2}$ as a response. Equivalently,
this contrast interprets ``had $A_2$ remained unchanged''
to mean ``we restrict attention to those
people whose value of $A_{2}$ would still have been $a_{2}$, even under an
intervention that set $A_{1}$ to a different value.''

Although the PSDE is an interesting parameter in many settings
(\cite*{gilbert:bosch:hudgens:biometrics:2003}), it has
drawbacks beyond the obvious (but perhaps less important) ones that
neither the
parameter itself nor the subgroup conditioned on are nonparametrically
identified. In fact, having just defined the PSDE parameter,
\citeauthor{robins:1986} (\citeyear{robins:1986})
criticized it for its lack of transitivity when there is
a non-null direct effect of $A_1$ and $A_{1}$ has more than two levels;
that is, for a given $a_2$, the
PSDEs comparing $a_{1}=0$ with $a_{1}=1$ and $a_{1}=1$ with $a_{1}=2$ may
both be positive but the PSDE comparing $a_{1}=0$ with $a_{1}=2$ may be
negative. \citeauthor{Robi:Rotn:Vans:disc:2007} (\citeyear{Robi:Rotn:Vans:disc:2007})
noted that the PSDE is
undefined when $A_{1}$ has an effect on every subject's $A_{2}$,
a situation that can easily occur if $A_2$ is continuous.
In that event, a natural strategy would be
to, say, dichotomize $A_{2}$. However, %
\citeauthor{Robi:Rotn:Vans:disc:2007} (\citeyear{Robi:Rotn:Vans:disc:2007})
showed that the PSDE in principal
stratum $a_{2}^{\ast}$ of the dichotomized variable may fail to retain any
meaningful substantive interpretation.

\subsection{\texorpdfstring{Pure Direct Effects (PDE)\protect\footnote{%
\citeauthor{pearl:indirect:01} (\citeyear{pearl:indirect:01}) adopted the definition given by %
\citet{Robi:Gree:iden:1992} 
 but changed nomenclature. He refers to the pure
direct effect as a ``natural'' direct effect.}}{Pure Direct Effects (PDE)}}\label{sec:pde}

Once it has been established that a treatment $A_{1}$ has a causal
effect on
a response $Y$, it is natural to ask what ``fraction'' of a the total effect
may be attributed to a given causal pathway. As an example, consider a RCT
in nonhypertensive smokers of the effect of an anti-smoking
intervention ($A_{1}$) on the outcome myocardial infarction (MI) at 2 years ($Y$). For
simplicity, assume everyone in the intervention arm and no one in the
placebo arm quit cigarettes, that all subjects were tested for new-onset
hypertension $A_{2}$ at the end of the first year, and no subject suffered
an MI in the first year. Hence, $A_{1}$, $A_{2}$ and $Y$ occur in that
order. Suppose the trial showed smoking cessation had a beneficial
effect on
both hypertension and MI. It is natural to consider the query:
``What fraction of the total effect of smoking cessation
$A_{1}$ on MI $Y$ is through a pathway that does not involve
hypertension $A_{2}$?''

\citet{Robi:Gree:iden:1992}
formalized this question via the following
counterfactual contrast, which they termed the ``pure direct effect'':
\[
Y\bigl\{a_1 = 1,A_2(a_1 = 0)\bigr\}-Y
\bigl\{a_1 = 0,A_2(a_1 = 0)\bigr\}.
\]

The second term here is simply $Y(a_{1} = 0)$.\footnote{%
This follows by consistency.} The contrast is thus the difference
between two quantities: first, the
outcome $Y$ that would result if we set $a_{1}$ to $1$, while ``holding
fixed'' $a_{2}$ at the value $A_{2}(a_{1} = 0)$ that it would have taken
had $a_{1}$ been $0$; second, the outcome $Y$ that would result from simply
setting $a_{1}$
to $0$ [and thus having $A_{2}$ again take the value $A_{2}(a_{1} = 0)$].
Thus, the Pure Direct Effect interprets had ``$A_{2}$
remained unchanged'' to mean ``had (somehow) $A_{2}$ taken the
value that it would have taken had we fixed $A_{1}$ to~$0$.''
The contrast thus represents the effect of
$A_{1}$ on $Y$ had the effect of $A_{1}$ on hypertension $A_{2}$ been
blocked. As for the CDE, to be well-defined, potential outcomes
$Y(a_{1},a_{2})$ must be well-defined. As a summary measure of the direct
effect of (a binary variable) $A_{1}$ on $Y$, the PDE has
the advantage (relative to the CDE and PSDE) that it is a single number.

The average pure direct effect is defined as\footnote{%
\citeauthor{Robi:Gree:iden:1992} (\citeyear{Robi:Gree:iden:1992})
also defined the total indirect effect (TIE)
of $A_{1}$ on $Y$ through $A_{2}$ to be
\[
E\bigl[ Y\bigl\{a_{1} = 1,A_{2}(a_{1} = 1)\bigr
\}\bigr] - E\bigl[Y\bigl\{a_{1} = 1,A_{2}(a_{1} =
0)\bigr\}\bigr] .
\]
It follows that the total effect $E[ Y\{a_{1} = 1\}]
-E[Y\{a_{1} = 0\}] $ can then be decomposed as the sum of the PDE and
the TIE.}
\begin{eqnarray*}
\operatorname{PDE} &=& E\bigl[ Y\bigl\{a_{1} = 1,A_{2}(a_{1}
= 0)\bigr\}\bigr]
\\
&&{} -E\bigl[Y\bigl(a_{1} = 0,A_{2}(a_{1} = 0)
\bigr)\bigr] .
\end{eqnarray*}
Thus, the ratio of the PDE to the total effect
$E[ Y\{a_{1} = 1\}] -E[Y\{a_{1} = 0\}] $ is the
fraction of the total that is through a pathway
that does not involve hypertension ($A_2$).

Unlike the PSDE, the PDE is an average over the full population. However,
unlike the CDE, the PDE is not nonparametrically identified under the
FFRCISTG model associated with the simple DAG shown in
Figure \ref{fig:no-confound}(a).
\citeauthor{robins:mcm:2011} (\citeyear{robins:mcm:2011}, App.~C)
computed bounds for the PDE under the
FFRCISTG associated with this DAG.

\citeauthor{pearl:indirect:01} (\citeyear{pearl:indirect:01})
obtains identification of the PDE
under the DAG in Figure~\ref{fig:no-confound}(a) by imposing stronger
counterfactual independence assumptions, via a Nonparametric Structural
Equation Model with Independent Errors (NPSEM-IE).\footnote{%
In more detail, the FFRCISTG associated with
Figures~\ref{fig:no-confound}(a) and (b)
assumes for all $a_{1}$, $a_{2}$,
%
\begin{equation}\label{eq:ffrcistgforpde}
\quad Y(a_{1},a_{2}),A_{2}(a_{1})
\Perp A_{1},\quad Y(a_{1},a_{2}) \Perp
A_{2}(a_{1})\mid A_{1},
\end{equation}
which may be read directly from the SWIG shown in
Figure~\ref{fig:no-confound}(b); recall that red nodes are always blocked when applying
d-separation. In contrast, Pearl's NPSEM-IE also implies the
independence
%
\begin{equation}
\label{eq:npsem-ie} Y(a_{1},a_{2}) \Perp A_{2}
\bigl(a_{1}^{*}\bigr)\mid A_{1},
\end{equation}
when $a_1\neq a_1^*$. Independence (\ref{eq:npsem-ie}), which is needed
in order for the PDE to be identified, is a ``cross-world'' independence since
$Y(a_{1},a_{2})$ and $A_{2}(a_{1}^{*})$ could never (even in principle)
both be observed in any randomized experiment.}
Under these assumptions,
\citeauthor{pearl:indirect:01} (\citeyear{pearl:indirect:01}) obtains the
following identifying formula:
%
\begin{eqnarray}\label{eq:mediation}
&& \sum_{a_{2}} \bigl\{ E[Y\mid A_{1} = 1,A_{2} = a_{2}]
\nonumber\\
&&\hspace*{19pt}{}
-E[Y\mid A_{1} = 0,A_{2} = a_{2}] \bigr\}
\\
&&\hspace*{14pt}{} \cdot
P(A_{2} = a_{2}\mid A_{1} = 0),
\nonumber
\end{eqnarray}
which he calls the ``Mediation Formula.''

\citeauthor{robins:mcm:2011} (\citeyear{robins:mcm:2011}) noted that
the additional assumptions made by the NPSEM-IE are not
testable, even in principle, via a randomized experiment. Consequently, this
formula represents a departure from the principle, originating with %
\citeauthor{neyman:sur:1923} (\citeyear{neyman:sur:1923}),
that causation be reducible to experimental
interventions, often expressed in the slogan ``no causation without
manipulation.''\footnote{%
A point freely acknowledged by
\citeauthor{pearl:myths:2012} (\citeyear{pearl:myths:2012}) who argues that
causation should be viewed as more primitive than intervention.} 
\citeauthor{robins:mcm:2011} (\citeyear{robins:mcm:2011})
achieve a rapprochement between these opposing
positions by showing that the formula (\ref{eq:mediation}) is equal
to the g-formula associated with an intervention on two treatment variables
not appearing on the graph (but having deterministic relations with $A_{1}$)
under the assumption that one of the variables has no direct effect on $A_{2}
$ and the other has no direct effect on $Y$. Hence, under this
assumption and
in the absence of confounding, the effect of this intervention on $Y$ is
point identified by (\ref{eq:mediation}).\footnote{%
This point identification is not a ``free lunch'':
\citeauthor{robins:mcm:2011} (\citeyear{robins:mcm:2011})
show that it is these additional assumptions
that have reduced the FFRCISTG bounds for the PDE to a point.
This is a consequence of the fact that these assumptions induce a model
\textit{for the original variables $\{A_1, A_2(a_1), Y(a_1,a_2)\}$}
that is a strict submodel of the original FFRCISTG model.

Hence to justify applying the mediation formula by this route one must first
be able to specify in detail the additional treatment variables and the
associated intervention
so as to make the relevant potential outcomes well-defined. In
addition, one must
be able to argue on substantive grounds for the plausibility of the required
no direct effect assumptions and deterministic relations.

It should also be noted that even
under Pearl's NPSEM-IE model the PDE is not identified in causal graphs,
such as those in Figures~\ref{fig:seq-rand} and \ref{fig:seq-rand-variant}
that contain a variable (whether observed or unobserved) that is
present both on a directed
pathway from $A_1$ to $A_2$ and on a pathway from $A_1$ to $Y$.}

Although there was a literature on direct effects
in linear structural equation models
(see, e.g., \cite*{blalock1971causal})
that preceded \citeauthor{robins:1986} (\citeyear{robins:1986}) and
\citeauthor{Robi:Gree:iden:1992} (\citeyear{Robi:Gree:iden:1992}),
the distinction between the CDE and PDE did not arise since in linear models
these notions are equivalent.\footnote{Note that in a linear
structural equation model
the PSDE is not defined unless $A_1$ has no effect on $A_2$.}

\begin{figure*}

\includegraphics{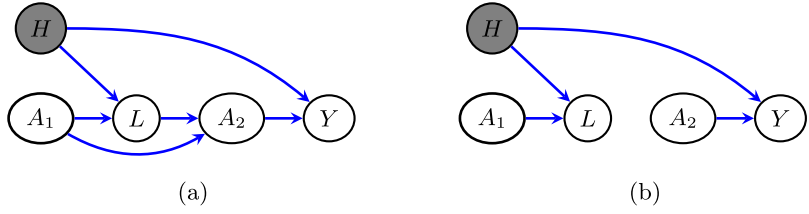}

\caption{\textup{(a)} A DAG representing the sequentially randomized experiment shown
in Figure \protect\ref{fig:seq-rand} but where there is no direct
effect of $A_1$ on $Y$ relative to $A_2$;
\textup{(b)} a DAG representing the pseudo-population
obtained by re-weighting the distribution with weights proportional to
$1/f(A_2 \mid L, A_1)$.}
\label{fig:seq-rand2}
\end{figure*}

\subsection{The Direct Effect Null}\label{sec:direct-null}

\citeauthor{robins:1986} (\citeyear{robins:1986}, Section~8)
considered the null hypothesis that
$Y(a_{1},a_{2})
$ does not depend on $a_{1}$ for all $a_{2}$, which we term the
\textit{sharp null-hypothesis of no direct effect of $A_{1}$ on $Y$}
(\textit{relative to $A_{2}$}) or more simply as the ``sharp direct effect null.''

In the context of our running example with data
$ (A_{1},L,A_{2},Y )$, under (\ref{eq:statrand}) the sharp direct
effect null
implies the following constraint on the observed data distribution:
%
\begin{equation}
\label{eq:verma-constraint} \quad f_{a_{1},a_{2}}^{\ast}(y) \quad\mbox{is not a
function of } a_{1} \mbox{ for all }a_{2}.
\end{equation}
\citeauthor{robins:1986} (\citeyear{robins:1986}, Sections~8 and 9)
noted that this constraint (\ref{eq:verma-constraint})
is \emph{not} a conditional independence.
This is in contrast
to the $g$-null hypothesis which we have seen is equivalent to the
independencies in (ii) of Section~\ref{sec:snm} [when
equation~(\ref{eq:indg}) holds for all $g\in\mathbb{G}$].\footnote{
Results in \citeauthor{pearl95on} (\citeyear{pearl95on})
imply that under the sharp direct effect
null the FFRCISTGs associated with the DAGs
shown in Figures~\ref{fig:seq-rand} and \ref{fig:seq-rand-variant} also
imply inequality restrictions similar to Bell's inequality in Quantum
Mechanics. See \citeauthor{gill:2014} (\citeyear{gill:2014}) for discussion of statistical issues
arising from experimental tests of Bell's inequality.}
He concluded that, in contrast to the $g$-null hypothesis, the
constraint (\ref{eq:verma-constraint}), and thus the sharp direct
effect null, cannot be
tested using case control data with unknown case and control sampling
fractions.\footnote{%
To our knowledge, it is the first such causal null hypothesis
considered in Epidemiology for which this is the case.}
This constraint~(\ref{eq:verma-constraint}) was later independently
discovered by
\citeauthor{verma:pearl:equivalence:1990} (\citeyear{verma:pearl:equivalence:1990})
and for this reason is
called the ``Verma constraint'' in the Computer Science literature.%

\citeauthor{robins:1999} (\citeyear{robins:1999}) noted that, though
(\ref{eq:verma-constraint}) is not a conditional independence in the observed
data distribution, it does correspond to a conditional independence,
but in a weighted distribution with weights proportional to
$1/f(A_{2}\mid A_{1},L)$.\footnote{This observation motivated the development of
graphical ``nested'' Markov models
that encode constraints such as (\ref{eq:verma-constraint}) in addition to
ordinary conditional independence relations; see
the discussion of ``Causal Discovery'' in Section~\ref{sec:otherwork} below.}
This can be understood from the informal discussion following
equation~(\ref{eq:msm2}) in the previous
section: there it was noted that given the
FFRCISTG corresponding to the DAG in Figure~\ref{fig:seq-rand}, reweighting
by $1/f(A_{2}\mid A_{1},L)$ corresponds to removing both edges into $A_{2}$.
Hence, if the edges $A_{1}\rightarrow Y$ and $L\rightarrow Y$ are not
present, so that the sharp direct effect null holds, as in Figure
\ref{fig:seq-rand2}(a), then the reweighted population is described by the DAG
in Figure~\ref{fig:seq-rand2}(b). It then follows from the d-separation
relations on this DAG that
$Y \Perp  A_{1}\mid A_{2}$ in the reweighted distribution.

This fact can also be seen as follows. If, in our running example
from Section~\ref{sec:tree-graph},
$A_{1}$, $A_{2}$, $Y$ are all binary, the sharp direct effect null implies
that $\beta_{1}^{\ast}=\beta_{3}^{\ast}=0$ in the saturated MSM with
\[
\Phi^{-1}\bigl\{E\bigl[Y(a_{1},a_{2})\bigr]\bigr
\}=\beta_{0}^{\ast} + \beta_{1}^{\ast}a_{1}+
\beta_{2}^{\ast}a_{2}+\beta_{3}^{\ast}a_{1}a_{2}.
\]
Since $\beta_{1}^{\ast}$ and $\beta_{3}^{\ast}$ are the associational
parameters of the weighted distribution, their being zero implies the
conditional independence $Y \Perp  A_{1}\mid A_{2}$ under this
weighted distribution.

In more complex longitudinal settings,
with the number of treatment times $k$ exceeding $2$,
all the parameters multiplying terms
containing a particular treatment variable in a MSM may be zero, yet there
may still be evidence in the data that the sharp direct effect null for that
variable is false. This is directly analogous to the limitation
of MSMs relative to SNMs
with regard to the sharp null hypothesis (\ref{g-null}) of no effect of
any treatment that we noted at the end of Section~\ref{sec:msm}.
To overcome this problem,
\citeauthor{robins:1999} (\citeyear{robins:1999})
introduced direct effect
structural nested models. In these models, which involve treatment at
$k$ time points, if all parameters multiplying a given $a_j$ take
the value $0$, then we can conclude that the distribution of the observables
do not
refute the natural extension of (\ref{eq:verma-constraint}) to $k$
times. The latter is
implied by the sharp direct effect null that $a_j$ has no direct effect
on $Y$ holding
$a_{j+1},\ldots,a_k$ fixed.

\section{The Foundations of Statistics and Bayesian Inference}

\citet{Robins:Ritov:toward:1997} and
\citet{Robi:Wass:cond:2000} recognized
that the lack of robustness of estimators based on the g-formula in a
sequential randomized trial with known randomization probabilities had
implications for the foundations of statistics and for Bayesian inference.
To make their argument transparent, we will assume in our running example
(from Section~\ref{sec:tree-graph}) that
the density of $L$ is known and that $A_{1}=1$ with probability $1$
(hence we
drop $A_{1}$ from the notation). We will further assume the observed
data are $n$
i.i.d. copies of a random vector $ ( L,A_{2},Y ) $ with
$A_{2}$ and $Y$ binary and $L$ a $d\times1$ continuous vector with support on
the unit cube $ ( 0,1 )^{d}$. We consider a model for the law
of $ ( L,A_{2},Y )$ that assumes that the density $f^{\ast}( l ) $
of $L$ is known, that the treatment probability
$\pi^{\ast}( l)\equiv\Pr( A_{2}=1 \mid  L=l)$ lies in the
interval $ ( c,1-c ) $ for
some known $c>0$ and that
$b^{\ast}( l,a_{2}) \equiv E[Y \mid  L=l,A_{2}=a_{2}] $ is continuous in $l$.
Under this model, the likelihood function is
%
\begin{equation}
\mathcal{L}( b, \pi) = \mathcal{L}_{1}( b) \mathcal{L}_{2}(
\pi) ,
\end{equation}
where
%
\begin{eqnarray}
\mathcal{L}_{1}( b) &=& \prod_{i=1}^{n}f^{\ast}(
L_{i}) b(L_{i},A_{2,i}) ^{Y}
\nonumber
\\[-8pt]
\\[-8pt]
&&\hspace*{14pt} {} \cdot \bigl\{ 1-b( L_{i},A_{2,i}) \bigr
\} ^{1-Y},
\nonumber
\\
\qquad \mathcal{L}_{2}( \pi) &=& \prod_{i=1}^{n}
\pi_{2}( L_{i})^{A_{2,i}} \bigl\{ 1-\pi_{2}(
L_{i}) \bigr\} ^{1-A_{2,i}},
\end{eqnarray}
and $ ( b,\pi ) \in\mathcal{B}\times\bolds{\Pi}$. Here
$\mathcal{B}$ is the set of continuous functions from
$ ( 0,1 )^{d}\times \{ 0,1 \} $ to
$ ( 0,1 ) $ and $\bolds{\Pi}$
is the set of functions from $ ( 0,1 ) ^{d}$ to $ (
c,1-c ) $.

We assume the goal is inference about $\mu( b) $ where
$\mu( b) =\int b(l,1) f^{\ast}( l) \,{dl}$.
Under randomization, that is (\ref{eq:ind1})
and (\ref{eq:ind2}), $\mu( b^{\ast})$ is the counterfactual mean of $Y$ when
treatment is given at both times.

When $\pi^{\ast}$ is unknown,
\citeauthor{Robins:Ritov:toward:1997} (\citeyear{Robins:Ritov:toward:1997}) showed that
no estimator of $\mu( b^{\ast}) $ exists that is uniformly consistent over
all $\mathcal{B}\times\bolds{\Pi}$. They also showed that even if
$\pi^{\ast}$ is known, any estimator that does not use knowledge of
$\pi^{\ast}$ cannot be uniformly consistent over
$\mathcal{B}\times \{ \pi^{\ast} \} $ for all
$\pi^{\ast}$. However, there do exist estimators that
depend on $\pi^{\ast}$ that are uniformly $\sqrt{n} $-consistent for
$\mu(b^{\ast}) $ over $\mathcal{B}\times \{ \pi^{\ast} \} $ for
all $\pi^{\ast}$. The Horvitz--Thompson estimator
$\mathbb{P}_{n}\{A_{2}Y/\pi^{\ast}( L) \} $ is a simple example.

\citeauthor{Robins:Ritov:toward:1997} (\citeyear{Robins:Ritov:toward:1997})
concluded that, in this example, any method
of estimation that obeys the likelihood principle such as maximum likelihood
or Bayesian estimation with independent priors on $b$ and~$\pi$, must fail
to be uniformly consistent. This is because any procedure that obeys the
likelihood principle must result in the same inference for $\mu(b^{\ast})$
regardless of $\pi^{\ast}$, even when $\pi^{\ast}$ becomes known. %
\citeauthor{Robi:Wass:cond:2000} (\citeyear{Robi:Wass:cond:2000})
noted that this example illustrates that the
likelihood principle and frequentist performance can be in severe conflict
in that any procedure with good frequentist properties must violate the
likelihood principle.\footnote{%
In response \citeauthor{robins:optimal:2004} (\citeyear{robins:optimal:2004}, Section~5.2)
offered a Bayes--frequentist compromise that combines honest subjective Bayesian
decision making under uncertainty with good frequentist behavior even when,
as above, the model is so large and the likelihood function so complex that
standard (uncompromised) Bayes procedures have poor frequentist performance.
The key to the compromise is that the Bayesian decision maker is only
allowed to observe a specified vector function of $X$ [depending on the
known $\pi^{\ast}( X) $] but not $X$ itself.} 
\citeauthor{ritov:2014} (\citeyear{ritov:2014})
in this volume extends this discussion in many directions.

\section{Semiparametric Efficiency and Double Robustness in Missing
Data and Causal Inference Models}\label{sec:semipar-eff}

\citeauthor{robins:rotnitzky:recovery:1992} (\citeyear{robins:rotnitzky:recovery:1992})
recognized that the inferential
problem of estimation of the mean $E[ Y(g)] $ (when identified by
the g-formula) of a response $Y$ under a regime $g$ is a special case of
the \textit{general problem} of estimating the parameters of an arbitrary
semi-parametric model in the presence of data that had been coarsened at
random (\cite*{Heitjan:Rubin:1991}).\footnote{%
Given complete data $X$, an always observed coarsening variable~$R$,
and a
known coarsening function $x_{(r)}=c(r,x)$, \emph{coarsening at random}
(CAR) is said to hold if $\Pr(R=r \mid  X)$ depends only on
$X_{(r)}$, the observed data part of $X$. %
\citeauthor{robins:rotnitzky:recovery:1992} (\citeyear{robins:rotnitzky:recovery:1992}),
\citeauthor{Gill:van:Robi:coar:1997} (\citeyear{Gill:van:Robi:coar:1997})
and %
\citeauthor{cator2004} (\citeyear{cator2004})
showed that in certain models assuming CAR places no
restrictions on the distribution of the observed data. For such models, we
can pretend CAR holds when our goal is estimation of functionals of the
observed data distribution. This trick often helps to derive efficient
estimators of the functional. In this section, we assume that the
distribution of the observables is compatible with CAR, and further,
that in
the estimation problems that we consider, CAR may be assumed to hold without
loss of generality.

In fact, this
is the case in the context of our running causal inference example from
Section~\ref{sec:tree-graph}.
Specifically, let $X= \{ Y(a_{1},a_{2}),L(a_{1});a_{j}\in \{
0,1 \} ,j=1,2 \} $, $R= ( A_{1},A_{2} ) $, and
$X_{ (a_{1},a_{2} ) }= \{ Y(a_{1},a_{2}),L(a_{1}) \} $.
Consider a model $M_{X}$ for $X$ that specifies
(i) $ \{ Y(1,a_{2}),L(1);a_{2}\in \{ 0,1 \}  \} \Perp
 \{ Y(0,a_{2}),L(0);a_{2}\in \{ 0,1 \}  \}$ and (ii)
$Y(a_{1},1) \Perp Y(a_{1},0) \mid  L(a_{1})$ for $a_{1}\in \{ 0,1
\}$. Results in \citeauthor{gill2001} (\citeyear{gill2001}, Section~6) and
\citeauthor{robins00marginal} (\citeyear{robins00marginal}, Sections~2.1 and 4.2)
show that (a) model $M_{X}$ places no further restrictions
on the distribution of the observed data $ ( A_{1},A_{2},L,Y )
= ( A_{1},A_{2},L( A_{1}),Y(A_{1},A_{2}) )$, (b)
given model $M_{X}$, the additional independences
$X \Perp A_{1}$ and $X \Perp A_{2} \mid  A_{1},L$ together also place no further restrictions on the
distribution of the
observed data $ ( A_{1},A_{2},L,Y ) $ and are equivalent to assuming
CAR. Further, the independences in (b) imply (\ref{eq:indg}) so that
$f_{Y(g)}(y)$ is identified by the g-formula $f_{g}^{\ast}(y)$.}
%

This viewpoint led them to recognize that the IPCW and IPTW estimators
described earlier were not fully efficient. To obtain efficient
estimators, %
\citet{robins:rotnitzky:recovery:1992} and
\citet{robins:rotnitzky:zhao:1994}
used the theory of semiparametric efficiency bounds %
(\cite*{bickel:klaasen:ritov:wellner:1993}; \cite*{van:on:1991}) to derive
representations for the efficient score, the efficient influence function,
the semiparametric variance bound, and the influence function of any
asymptotically linear estimator in this \textit{general} problem. The books
by \citeauthor{tsiatis:2006} (\citeyear{tsiatis:2006}) and by
\citeauthor{vdL:robins:2003} (\citeyear{vdL:robins:2003}) provide thorough
treatments. The generality of these results allowed Robins and his principal
collaborators Mark van der Laan and Andrea Rotnitzky to solve many open
problems in the analysis of semiparametric models. For example, they
used the
efficient score representation theorem to derive locally efficient
semiparametric estimators in many models of importance in biostatistics.
Some examples include conditional mean models with missing regressors and/or
responses (\cite*{robins:rotnitzky:zhao:1994}; \cite*{Rotn:Robi:semi:1995}),
bivariate survival (\cite*{quale2006locally}) and
multivariate survival models with explainable dependent censoring
(\cite*{van2002locally}).\footnote{More recently, in the context of a RCT,
\citet{tsiatis2008covariate} and
\citet{moore2009covariate},
following the strategy of
\citet{robins:rotnitzky:recovery:1992},
studied variants of the locally efficient tests and estimators of
\citet{Scha:Rotn:Robi:adju:1999}
to increase efficiency and power by utilizing data on covariates.}

In coarsened at random data models, whether missing data or causal inference
models, locally efficient semiparametric estimators are also doubly
robust %
(\cite*{Scha:Rotn:Robi:adju:1999}, pages 1141--1144) and
(\cite*{Robins:Rotnitzky:comment:on:bickel:2001}). See the book %
(\cite*{vdL:robins:2003}) for details and for many examples of doubly robust
estimators. Doubly robust estimators had been discovered earlier in special
cases. In fact, \citeauthor{Firth:Bennet:1998} (\citeyear{Firth:Bennet:1998})
note that the so-called model-assisted
regression estimator of a finite population mean %
of \citeauthor{Cass:Srnd:Wret:some:1976} (\citeyear{Cass:Srnd:Wret:some:1976})
is design consistent which is tantamount to
being doubly robust. See
\citeauthor{Robins:Rotnitzky:comment:on:bickel:2001} (\citeyear{Robins:Rotnitzky:comment:on:bickel:2001})
for other precursors.

In the context of our running example, from Section~\ref{sec:tree-graph},
suppose (\ref{eq:statrand}) holds. An
estimator $\widehat{\mu}_{\mathrm{dr}}$ of
$\mu=E[ Y( a_{1},a_{2})]=f_{a_{1},a_{2}}^{\ast}(1)$
for, say $a_{1}=a_{2}=1$, is said to be
\textit{doubly robust} (DR) if it is consistent when either (i) a model
for $\pi(L) \equiv\Pr( A_{2}=1 \mid  A_{1}=1,L) $ or (ii)~a model for
$b( L) \equiv E[ Y \mid  A_{1}=1,L,A_{2}=1]$ is correct.
When $L$ is high dimensional and, as in an observational study,
$\pi(\cdot)$ is unknown,
double robustness is a desirable property because
model misspecification is generally unavoidable, even when we use flexible,
high dimensional, semiparametric models in (i) and (ii). In fact, DR
estimators have advantages even when, as is usually the case, the
models in
(i) and (ii) are both incorrect. This happens because the bias of the DR
estimator $\widehat{\mu}_{\mathrm{dr}}$ is of second order, and thus
generally less
than the bias of a non-DR estimator (such as a standard IPTW
estimator). By
second order, we mean that the bias of $\widehat{\mu}_{\mathrm{dr}}$ depends on the
product of the error made in the estimation of
$\Pr(A_{2}=1 \mid  A_{1}=1,L) $ times the error made in the estimation of
$E[ Y \mid  A_{1}=1,L,A_{2}=1] $.

\citeauthor{Scha:Rotn:Robi:adju:1999} (\citeyear{Scha:Rotn:Robi:adju:1999})
noted that the locally
efficient estimator of
\citet{robins:rotnitzky:zhao:1994}
\begin{eqnarray*}
\widetilde{\mu}_{\mathrm{dr}}
&=& \bigl\{ \mathbb{P}_{n}[A_{1}] \bigr\}^{-1}
\\
&&{}\cdot \mathbb{P}_{n} \biggl[A_{1} \biggl\{ \frac{A_{2}}{\widehat{\pi}(L) }Y
- \biggl\{ \frac{A_{2}}{\widehat{\pi}( L) }-1
\biggr\} \widehat{b}( L) \biggr\} \biggr]
\end{eqnarray*}
is doubly robust where $\widehat{\pi}( L) $ and $\widehat{b}( L) $
are estimators of $\pi( L) $ and $b(L)$.
Unfortunately, in finite samples this estimator may fail to lie
in the parameter space for $\mu$, that is,~the interval $[0,1]$ if $Y$ is
binary. In response, \citeauthor{Scha:Rotn:Robi:adju:1999} (\citeyear{Scha:Rotn:Robi:adju:1999})
 proposed a plug-in DR
estimator, the doubly robust regression estimator
\[
\widehat{\mu}_{\mathrm{dr},\mathrm{reg}} = \bigl\{ \mathbb{P}_{n} [
A_{1} ] \bigr\}^{-1}\mathbb{P}_{n} \bigl\{
A_{1} \widehat{b}( L) \bigr\},
\]
where now $\widehat{b}( L) =\operatorname{expit}\{ m( L;\widehat{\eta})
+ \widehat{\theta}/\widehat{\pi}( L)\}$ and
$( \widehat{\eta},\widehat{\theta}) $ are obtained by
fitting by maximum likelihood the logistic regression model
$\Pr(Y=1 \mid  A_{1}=1,\allowbreak L,A_{2}=1) =\operatorname{expit}\{ m( L;\eta) +\theta
/\widehat{\pi}( L) \} $ to subjects with
$A_{1}=1$, $A_{2}=1$. Here, $m( L;\eta)$ is a user-specified function of $L$
and of the Euclidean parameter $\eta$.

\citeauthor{Robi:robust:1999} (\citeyear{Robi:robust:1999}) and
\citeauthor{Bang:Robi:doub:2005} (\citeyear{Bang:Robi:doub:2005})
obtained plug-in DR regression estimators in longitudinal
missing data and causal inference models by reexpressing the g-formula
as a
sequence of iterated conditional expectations.

\citeauthor{van:Rubi:targ:2006} (\citeyear{van:Rubi:targ:2006})
proposed a clever general method for obtaining
plug-in DR estimators called targeted maximum likelihood. In our setting,
the method yields an estimator $\widehat{\mu}_{\mathrm{dr},\mathrm{TMLE}}$ that differs
from $\widehat{\mu}_{\mathrm{dr},\mathrm{reg}}$ only in that $\widehat{b}( L) $ is now
given by $\operatorname{expit}\{ \widehat{m}( L) +\widehat{\theta}_{\mathrm{greedy}}/\widehat{\pi}( L) \} $
where $\widehat{\theta}_{\mathrm{greedy}}$ is again obtained by maximum likelihood but with a fixed
offset $\widehat{m}( L) $. This offset is an estimator of
$\Pr(Y=1 \mid  A_{1}=1,L,A_{2}=1) $
that might be
obtained using flexible machine learning
methods. Similar comments apply to models considered by Bang and Robins (\citeyear{Bang:Robi:doub:2005}). Since 2006 there has been an explosion of research that has
produced doubly robust estimators with much improved large sample efficiency
and finite sample performance;
\citeauthor{rotnitkzy:vansteelandt:2014} (\citeyear{rotnitkzy:vansteelandt:2014})
give a review.

We note that CAR models are not the only models that admit doubly robust
estimators. For example,
\citeauthor{Scha:Rotn:Robi:adju:1999} (\citeyear{Scha:Rotn:Robi:adju:1999})
exhibited doubly robust estimators in models with nonignorable missingness. %
\citeauthor{Robins:Rotnitzky:comment:on:bickel:2001} (\citeyear{Robins:Rotnitzky:comment:on:bickel:2001})
derived sufficient
conditions, satisfied by many non-CAR models, that imply the existence of
doubly robust estimators. Recently, doubly robust estimators have been
obtained in a wide variety of models. See
\citeauthor{dudik:2014} (\citeyear{dudik:2014})
in this volume for an interesting example.

\section{Higher Order Influence Functions}

It may happen that the second-order bias of a doubly-robust estimator
$\widehat{\mu}_{\mathrm{dr}}$ decreases slower to 0 with $n$ than $n^{-1/2}$, and
thus the bias exceeds the standard error of the estimator. In that case,
confidence intervals for $\mu$ based on $\widehat{\mu}_{\mathrm{dr}}$ fail to cover
at their nominal rate even in large samples. Furthermore, in such a
case, in
terms of mean squared error, $\widehat{\mu}_{\mathrm{dr}}$ does not optimally trade
off bias and variance. In an attempt to address these problems, %
\citeauthor{robins:higher:2008} (\citeyear{robins:higher:2008})
developed a theory of point and interval
estimation based on higher order influence functions and use this
theory to
construct estimators of $\mu$ that improve on $\widehat{\mu}_{\mathrm{dr}}$.
Higher order influence functions are higher order U-statistics. The theory
of \citeauthor{robins:higher:2008} (\citeyear{robins:higher:2008})
extends to higher order the first order
semiparametric inference theory of
\citeauthor{bickel:klaasen:ritov:wellner:1993} (\citeyear{bickel:klaasen:ritov:wellner:1993})
and \citeauthor{van:on:1991} (\citeyear{van:on:1991}). In this issue,
\citeauthor{vandervaart:2014} (\citeyear{vandervaart:2014}) gives a
masterful review of this theory. Here, we present
an interesting result found in
\citeauthor{robins:higher:2008} (\citeyear{robins:higher:2008})
that can be understood in isolation
from the general theory and conclude with an open estimation problem.%

\citeauthor{robins:higher:2008} (\citeyear{robins:higher:2008})
consider the question of whether, for estimation of a conditional variance,
random regressors provide for faster rates of convergence than do fixed
regressors, and, if so, how? They consider a setting in which $n$ i.i.d.~copies
of $ ( Y,X ) $ are observed with $X$ a $d$-dimensional random
vector, with bounded density $f( \cdot) $
absolutely continuous w.r.t.~the uniform measure on the unit cube
$ (0,1 ) ^{d}$. The regression function
$b( \cdot) =E  [Y \mid X=\cdot ]$
is assumed to lie in a given H\"{o}lder ball with H\"{o}lder
exponent $\beta<1$.
\footnote{A function $b( \cdot) $ lies in the
H\"{o}lder ball $H(\beta,C)$ with H\"{o}lder exponent $\beta>0$ and radius
$C>0$, if and only if $b( \cdot) $ is bounded in supremum norm by
$C$ and all partial derivatives of $b(x)$ up to order
$ \lfloor\beta \rfloor$ exist, and all partial derivatives of order
$ \lfloor\beta \rfloor$ are Lipschitz with exponent
$ ( \beta- \lfloor\beta \rfloor )$ and constant $C$.}
The goal is to estimate $E[\hbox{Var}  \{ Y \mid  X \} ]$
under the homoscedastic semiparametric model
$\operatorname{Var}[ Y \mid  X] =\sigma^{2}$. Under this model, the authors
construct a simple estimator
$\widehat{\sigma}^{2}$ that converges at rate $n^{-\fraca{4\beta/d}{1+4\beta/d}}$,
when $\beta/d<1/4$.

\citeauthor{Wang:Brow:Cai:Levi:effe:2008} (\citeyear{Wang:Brow:Cai:Levi:effe:2008}) and
\citeauthor{Cai2009126} (\citeyear{Cai2009126})
earlier proved that if $X_{i},i=1,\ldots,n$, are
nonrandom but equally spaced in $ ( 0,1 )^{d}$, the minimax rate
of convergence for the estimation of $\sigma^{2}$ is $n^{-2\beta/d}$
(when $\beta/d<1/4$) which
is slower than $n^{-\fraca{4\beta/d}{1+4\beta/d}}$. Thus, randomness
in $X$
allows for improved convergence rates even though no smoothness assumptions
are made regarding $f( \cdot)$.

To explain how this happens, we describe the estimator of
\citeauthor{robins:higher:2008} (\citeyear{robins:higher:2008}). The
unit cube in $\mathbb{R}^{d}$ is divided into $k=k( n ) =n^{\gamma}$,
$\gamma>1$ identical subcubes each with edge length $k^{-1/d}$.
A simple probability calculation shows
that the number of subcubes containing at least two observations is
$O_{p}( n^{2}/k)$. One may estimate $\sigma^{2}$ in each such
subcube by $( Y_{i}-Y_{j})^{2}/2$.%
\footnote{If a subcube contains
more than two observations, two are selected randomly, without replacement.}
An estimator $\widehat{\sigma}^{2}$ of $\sigma^{2}$ may then be constructed
by simply averaging
the subcube-specific estimates $ ( Y_{i}-Y_{j} ) ^{2}/2$ over
all the sub-cubes with at least two observations.
The rate of convergence of the estimator is maximized at
$n^{-\fraca{4\beta/d}{1+4\beta/d}}$ by taking
$k=n^{\fracz{2}{1+4\beta/d}}$.%
\footnote{
Observe that $E  [  ( Y_{i}-Y_{j} ) ^{2}/2 \mid  X_{i},X_{j} ]
=\sigma^{2}+ \{ b( X_{i}) -b( X_{j}) \} ^{2}/2$,
$\llvert  b( X_{i}) -b( X_{j})\rrvert =O  (\llVert  X_{i}-X_{j}
\rrVert ^{\beta} )$ as $\beta<1$,
and $\llVert  X_{i}-X_{j}\rrVert =d^{1/2}O( k^{-1/d})$
when $X_{i}$ and $X_{j}$ are in the same subcube. It follows that the
estimator has variance of order $k/n^{2}$ and bias of order
$O(k^{-2\beta/d})$. Variance and the squared bias are equated by
solving $k/n^{2}=k^{-4\beta/d}$ which gives $k=n^{\fracz{2}{1+4\beta/d}}$.
}

\citeauthor{robins:higher:2008} (\citeyear{robins:higher:2008})
conclude that the random design estimator has better bias
control, and hence converges faster than the optimal equal-spaced fixed $X$
estimator, because the random design estimator exploits the
$O_{p}  (n^{2}/n^{\fracz{2}{1+4\beta/d}} ) $ random fluctuations for which
the $X$'s corresponding to two different observations are only a
distance of $O  (  \{ n^{\fracz{2}{1+4\beta/d}} \}^{-1/d} )$ apart.

\subsection*{An Open Problem\protect\footnote{
Robins has been trying to find an answer to this question without
success for a number of years. He suggested that it is now time for some
crowd-sourcing.}}

Consider again the above setting with random $X$.
Suppose that $\beta/d$ remains less than $1/4$ but now $\beta>1$. Does
there still exist an estimator of $\sigma^{2}$ that converges at
$n^{-\fraca{4\beta/d}{1+4\beta/d}}$? Analogy with other nonparametric
estimation problems would suggest the answer is ``yes,'' but the question
remains unsolved.%
\footnote{The estimator given above does not attain this rate when
$\beta>1$ because it fails to exploit the fact that $b(\cdot)$ is
differentiable. In the interest of simplicity, we have posed this as a
problem in variance estimation. However,
\citeauthor{robins:higher:2008} (\citeyear{robins:higher:2008})
show that the estimation of the variance is mathematically isomorphic
to the estimation of $\theta$ in the semi-parametric regression model
$E[Y \mid  A,X]=\theta A +h(X)$, where $A$ is a binary treatment. In the
absence of confounding, $\theta$ encodes the causal effect of the treatment.}



\section{Other Work}\label{sec:otherwork}

The available space precludes a complete treatment of all of the topics that
Robins has worked on. We provide a brief description of selected additional
topics and a guide to the literature.

\subsection*{Analyzing Observational Studies as Nested Randomized Trials}

\citeauthor{hernan2008observational} (\citeyear{hernan2008observational}) and
\citet{hernan2005discussion}
conceptualize and analyze
observational studies of a time varying treatment as a nested sequence of
individual RCTs trials run by nature. Their analysis is closely related to
g-estimation of SNM (discussed in Section~\ref{sec:snm}). The critical
difference is that in these papers  Robins and Hern\'{a}n do not specify a SNM to
coherently link the
trial-specific effect estimates. This has benefits in that it makes the
analysis easier and also more familiar to users without training in SNMs.
The downside is that, in principle, this lack of coherence can result in
different analysts recommending, as optimal, contradictory
interventions %
(\citeauthor{robins2007invited} \citeyear{robins2007invited}).

\subsection*{Adjustment for ``Reverse Causation''}

Consider an epidemiological study of a time-  dependent treatment (say
cigarette smoking) on time to a disease of interest, say clinical lung
cancer. In this setting, uncontrolled confounding by undetected preclinical
lung cancer (often referred to as ``reverse causation") is a serious
problem. %
\citeauthor{robins2008causal} (\citeyear{robins2008causal})
develops analytic methods that may still
provide an unconfounded effect estimate, provided that (i) all subjects with
preclinical disease severe enough to affect treatment (i.e., smoking
behavior) at a given time $t$ will have their disease clinically diagnosed
within the next~$x$, say $2$ years and (ii) based on subject matter knowledge
an upper bound, for example, $3$ years, on $x$ is known.

\subsection*{Causal Discovery}

\citeauthor{cps93} (\citeyear{cps93}) and
\citet{pearlverm:tark} proposed statistical
methods that allowed one to draw causal conclusions from
associational data. These methods assume an underlying
causal DAG (or equivalently an FFRCISTG). If the
DAG is incomplete, then such a model imposes
conditional independence relations on the associated joint distribution
(via d-separation).
\citet{cps93} and
\citeauthor{pearlverm:tark} (\citeyear{pearlverm:tark}) made the
additional assumption that {\em all} conditional independence
relations that hold in the distribution of the observables
are implied by the underlying causal graph, an assumption
termed ``stability'' by
\citeauthor{pearlverm:tark} (\citeyear{pearlverm:tark}),
and ``faithfulness'' by
\citet{cps93}. Under this assumption, the underlying
DAG may be identified up to a (``Markov'') equivalence class.
\citet{cps93} proposed two algorithms that recover such a class,
entitled ``PC'' and ``FCI.'' While the former presupposes that there are no
unobserved common
causes, the latter explicitly allows for this possibility.

\citeauthor{robins:impossible:1999} (\citeyear{robins:impossible:1999}) and
\citeauthor{robins:uniform:2003} (\citeyear{robins:uniform:2003}) pointed
out that
although these procedures were consistent they were not uniformly
consistent. More recent papers
(\cite*{kalisch:2007}; \cite*{Colo:Maat:Kali:Rich:lear:2012})
recover uniform consistency for these algorithms by imposing additional
assumptions.
\citeauthor{spirtes:2014} (\citeyear{spirtes:2014})
in this volume extend this work by
developing a variant of the PC Algorithm which is uniformly consistent
under weaker assumptions.

\citeauthor{shpitser12parameter} (\citeyear{shpitser12parameter,shpitser2014introduction}),
building on
\citet{tian02on} and
\citeauthor{robins:1999} (\citeyear{robins:1999})
develop a theory of \textit{nested Markov models}
that relate the structure of a causal DAG to
conditional independence relations that arise after re-weighting; see
Section~\ref{sec:direct-null}. This
theory, in combination with the
theory of graphical Markov models based on Acyclic Directed Mixed
Graphs (\cite*{richardson:2002};
\cite*{richardson:2003};
\cite*{wermuth:11};
\cite*{evans2014};
\cite*{sadeghi2014}),
will facilitate the construction of more powerful\footnote{But still
not uniformly consistent!}
causal discovery algorithms that could (potentially) reveal much more
information
regarding the structure of a DAG containing hidden variables than algorithms
(such as FCI) that solely use conditional independence.

\subsection*{Extrapolation and Transportability of Treatment Effects}

Quality longitudinal data is often only available in high resource settings.
An important question is when and how can such data be used to inform the
choice of treatment strategy in low resource settings. To help answer this
question, \citeauthor{robins2008estimation} (\citeyear{robins2008estimation})
studied the extrapolation of optimal
dynamic treatment strategies between two HIV infected patient
populations. The
authors considered the treatment strategies $g_x$, of the same form
as those defined in Section~\ref{sec:msm}, namely, ``start
anti-retroviral therapy the first time at which the measured CD4 count
falls below $x$.'' Given a utility measure $Y$, their
goal is to find the regime $g_{x_{\mathrm{opt}}}$ that maximizes $E[ Y(g_x)]$
in the second low-resource population when good longitudinal data
are available only in the first high-resource population. Due to differences
in resources, the frequency of CD4 testing in the first population is much
greater than in the second and, furthermore, for logistical and/or financial
reasons, the testing frequencies cannot be altered. In this setting, the
authors derived conditions under which data from the first population is
sufficient to identify $g_{x_{\mathrm{opt}}}$ and construct IPTW estimators of
$g_{x_{\mathrm{opt}}}$
under those conditions. A key finding is that owing to the differential
rates of testing, a necessary condition for identification is that CD4
testing has no direct causal effect on $Y$ not through anti-retroviral
therapy. In this issue,
\citeauthor{pearl:2014} (\citeyear{pearl:2014}) study the related
question of transportability between populations using graphical tools.

\subsection*{Interference, Interactions and Quantum Mechanics}

Within a counterfactual causal model,
\citeauthor{cox1958} (\citeyear{cox1958}) defined there to
be \textit{interference between
treatments} if the response of some subject depends not only on their treatment
but on that of others as well. On the other hand,
\citeauthor{Vand:Robi:mini:2009} (\citeyear{Vand:Robi:mini:2009}) defined
two binary treatments
$ ( a_{1},a_{2} ) $ to be \textit{causally interacting} to cause
a binary response $Y$ if for some unit $Y( 1,1) \neq Y( 1,0) =Y(0,1) $;
\citeauthor{Vand:epis:2010} (\citeyear{Vand:epis:2010}) defined
the interaction to be \textit{epistatic} if $Y( 1,1)
\neq Y( 1,0) =Y(0,1) =Y( 0,0)$.
VanderWeele with his collaborators has developed a very general theory of
empirical tests for causal interaction of different types
(\cite*{Vand:Robi:mini:2009};
\cite*{Vand:epis:2010}, \citeyear{Vand:suff:2010};
\cite*{vanderweele2012}).

\citeauthor{robins2012proof} (\citeyear{robins2012proof})
showed, perhaps surprisingly, that this theory could
be used to give a simple but novel proof of an important result in quantum
mechanics known as Bell's theorem. The proof was based on two insights: The
first was that the consequent of Bell's theorem could, by using the
Neyman causal model, be recast as the statement that there is interference
between a certain pair of treatments. The second was to recognize that
empirical tests for causal interaction can be reinterpreted as tests for
certain forms of interference between treatments, including the form needed
to prove Bell's theorem.
\citeauthor{vanderweele2012mapping} (\citeyear{vanderweele2012mapping}) used this latter
insight to show that existing empirical tests for causal interactions could
be used to test for interference and spillover effects in vaccine
trials and
in many other settings in which interference and spillover effects may be
present. The papers
\citeauthor{ogburn:2014} (\citeyear{ogburn:2014}) and
\citet{vanderweele:2014}
in this issue
contain further results on interference and spillover effects.

\subsection*{Multiple Imputation}

\citeauthor{wang1998large} (\citeyear{wang1998large}) and
\citeauthor{robins2000inference} (\citeyear{robins2000inference})
studied the statistical properties of the multiple imputation approach to
missing data (\cite*{rubin2004multiple}). They derived a variance estimator
that is consistent for the asymptotic variance of a multiple imputation
estimator even under misspecification and incompatibility of the imputation
and the (complete data) analysis model. They also characterized the large
sample bias of the variance estimator proposed by
\citeauthor{Rubi:mult:1978} (\citeyear{Rubi:mult:1978}).

\subsection*{Posterior Predictive Checks}

\citeauthor{robins2000asymptotic} (\citeyear{robins2000asymptotic})
studied the asymptotic null distributions of
the posterior predictive p-value of
\citeauthor{rubin1984bayesianly} (\citeyear{rubin1984bayesianly}) and %
\citeauthor{guttman1967use} (\citeyear{guttman1967use})
and of the conditional predictive and partial
posterior predictive p-values of
\citeauthor{bayarri2000p} (\citeyear{bayarri2000p}). They found the latter
two p-values to have an asymptotic uniform distribution; in contrast they
found that the posterior predictive p-value could be very conservative,
thereby diminishing its power to detect a misspecified model.
In response, Robins et al. derived an adjusted version of the posterior predictive  p-value that  was asymptotically uniform.

\subsection*{Sensitivity Analysis}

Understanding that epidemiologists will almost  never succeed in collecting
data on all covariates needed to fully prevent confounding by unmeasured
factors and/or nonignorable missing data, Robins with collaborators Daniel
Scharfstein and Andrea Rotnitzky developed methods for conducting
sensitivity analyses. See, for example,
\citet{Scha:Rotn:Robi:adju:1999},
\citet{robins2000sensitivity}
and \citeauthor{robins2002covariance} (\citeyear{robins2002covariance}, pages~319--321).
In this issue,
\citeauthor{richardson:hudgens:2014} (\citeyear{richardson:hudgens:2014})
describe methods for sensitivity
analysis and present several applied examples.

\subsection*{Public Health Impact}

Finally, we have not discussed the large impact of the methods that Robins
introduced on the substantive analysis of longitudinal data in epidemiology
and other fields.
Many researchers have been involved in transforming Robins' work on
time-varying treatments
into increasingly reliable, robust analytic tools and in applying these
tools to help answer questions of public health importance.

%

\section*{List of Acronyms Used}\label{acronyms}
\begin{center}
{\footnotesize
\begin{tabular}{@{}r@{\quad}l@{\quad}p{133pt}@{}}
CAR:                       & Section \ref{sec:semipar-eff}     & coarsened at random.                                                \\
CD4:                       & Section \ref{sec:tree-graph}      & (medical) cell line depleted by HIV.                                \\
CDE:                       & Section \ref{sec:cde}             & controlled direct effect.                                           \\
CMA:                       & Section \ref{sec:dags}            & causal Markov assumption.                                           \\
DAG:                       & Section \ref{sec:dags}            & directed acyclic graph.                                             \\
DR:                        & Section \ref{sec:semipar-eff}     & doubly robust.                                                      \\
dSWIG:                     & Section \ref{sec:dynamic-regimes} & dynamic single-world intervention graph.                            \\
FFRCISTG:                  & Section \ref{sec:tree-graph}      & finest fully randomized causally interpreted structured tree graph. \\
HIV:                       & Section \ref{sec:tree-graph}      & (medical) human immunodeficiency virus.                             \\
IPCW:                      & Section \ref{sec:censoring}       & inverse probability of censoring weighted.                          \\
IPTW:                      & Section \ref{sec:msm}             & inverse probability of treatment weighted.                          \\
ITT:                       & Section \ref{sec:censoring}       & intention to treat.                                                 \\
MI:                        & Section \ref{sec:pde}             & (medical) myocardial infarction.                                    \\
MSM:                       & Section \ref{sec:msm}             & marginal structural model.                                          \\
\end{tabular}}
\end{center}
\begin{center}
{\footnotesize
\begin{tabular}{@{}r@{\quad}l@{\quad}p{133pt}@{}}
NPSEM:                     & Section \ref{sec:tree-graph}      & nonparametric structural equation model.                            \\
NPSEM-IE:                  & Section \ref{sec:tree-graph}      & nonparametric structural equation model with independent errors.    \\
PDE:                       & Section \ref{sec:pde}             & pure direct effects.                                                \\
PSDE:                      & Section \ref{sec:psde}            & principal stratum direct effects.                                   \\
RCT:                       & Section \ref{sec:tree-graph}      & randomized clinical trial.                                          \\
SNM:                       & Section \ref{sec:snm}             & structural nested model.                                            \\
SNDM:                      & Section \ref{sec:snm}             & structural nested distribution model.                               \\
SNFTM:                     & Section \ref{sec:snm}             & structural nested failure time model.                               \\
SNMM:                      & Section \ref{sec:snm}             & structural nested mean model.                                       \\
SWIG:                      & Section \ref{sec:dags}            & single-world intervention graph.                                    \\
TIE:                       & Section \ref{sec:pde}             & total indirect effect.
\end{tabular}
}
\end{center}

\section*{Acknowledgments}

This work was supported by the US\ National Institutes of Health
Grant R01 AI032475.




\end{document}